\documentclass[aps,prd,
nofootinbib,
twocolumn,
showpacs
]{revtex4-1}
\usepackage{amsmath}
\usepackage{amssymb}
\usepackage{epsfig}
\usepackage{graphicx}
\usepackage{multirow}
\usepackage{color}
\usepackage[normal]{subfigure}
\usepackage{rotating}

\newcommand{\capdef}{}
\newcommand{\mycaption}[2][\capdef]{\renewcommand{\capdef}{#2}
\caption[#1]{{\footnotesize #2}}}

\newcommand{\bwt}{\begin{widetext}}
\newcommand{\ewt}{\end{widetext}}
\newcommand{\beq}{\begin{equation}}
\newcommand{\eeq}{\end{equation}}
\newcommand{\bdm}{\begin{displaymath}}
\newcommand{\edm}{\end{displaymath}}
\newcommand{\bea}{\begin{eqnarray}}
\newcommand{\eea}{\end{eqnarray}}
\newcommand{\nn}{\nonumber}

\def\eq#1{{Eq.~(\ref{#1})}}
\def\eqs#1#2{{Eqs.~(\ref{#1})--(\ref{#2})}}
\def\fig#1{{Fig.~\ref{#1}}}

\def\Table#1{{Table~\ref{#1}}}
\def\Tables#1#2{{Tables~\ref{#1}--\ref{#2}}}
\def\sect#1{{Sect.~\ref{#1}}}
\def\sects#1#2{{Sects.~\ref{#1}--\ref{#2}}}
\def\app#1{{Appendix~\ref{#1}}}
\def\apps#1#2{{Apps.~\ref{#1}--\ref{#2}}}
\def\vev#1{\left\langle #1\right\rangle}

\def\Tr{\mbox{Tr}\,}

\begin{document}

%%%%%%%%%%%%%%%%%%%%%%%%%%%%%%%%%%%%%%%%%%%%%%%%%%%%%%%%%%%%%%%%%%%%%%%
%%%%                       Title-page                              %%%%
%%%%%%%%%%%%%%%%%%%%%%%%%%%%%%%%%%%%%%%%%%%%%%%%%%%%%%%%%%%%%%%%%%%%%%%

%-------------------------------------------------------------------------------
\title{On the vacuum of the minimal nonsupersymmetric $SO(10)$ unification}
%-------------------------------------------------------------------------------
%\date{\today}
\date{December 8, 2009}
\author{Stefano Bertolini}\email{bertolin@sissa.it}
\author{Luca Di Luzio}\email{diluzio@sissa.it}
\affiliation{INFN, Sezione di Trieste, and SISSA,
Via Beirut 4, I-34014 Trieste, Italy}
\author{Michal Malinsk\'{y}}\email{malinsky@kth.se}
\affiliation{Theoretical Particle Physics Group,
Department of Theoretical Physics,
Royal Institute of Technology (KTH),
Roslagstullsbacken 21,
SE-106 91 Stockholm, Sweden.}
%------------------------------------------------------------------------------
\begin{abstract}
We study a class of nonsupersymmetric $SO(10)$ grand unified scenarios
where the first stage of the symmetry breaking is driven by the vacuum
expectation values
of the $45$-dimensional adjoint representation.
Three decade old results claim that such a Higgs setting may lead
exclusively to the flipped $SU(5)\otimes U(1)$ intermediate stage.
We show that this conclusion is actually an artifact of the tree level potential.
The study of the accidental global symmetries emerging
in various limits of the scalar potential offers a simple understanding
of the tree level result and a rationale for the drastic impact of
quantum corrections.
We scrutinize in detail the simplest and paradigmatic case of the
$45_{H}\oplus 16_{H}$ Higgs sector triggering the breaking
of $SO(10)$ to the standard electroweak model.
We show that the minimization of the
one-loop effective potential allows for intermediate
$SU(4)_C \otimes SU(2)_L \otimes U(1)_R$
and
$SU(3)_c \otimes SU(2)_L \otimes SU(2)_R \otimes U(1)_{B-L}$
symmetric stages as well.
These are the options favored by gauge unification.
Our results, that apply whenever the $SO(10)$ breaking is
triggered by $\vev{45_H}$, open
the path for hunting the simplest realistic scenario
of nonsupersymmetric $SO(10)$ grand unification.

%\vspace*{0ex}
\end{abstract}
%-------------------------------------------------------------------------------
%-------------------------------------------------------------------------------
\maketitle
% the footnote symbols are only redefined for the title page !
\renewcommand{\thefootnote}{\alph{footnote}}
\renewcommand{\thefootnote}{\fnsymbol{footnote}}
\renewcommand{\thefootnote}{\it\alph{footnote}}
\renewcommand{\thefootnote}{\arabic{footnote}}
\setcounter{footnote}{0}

%%%%%%%%%%%%%%%%%%%%%%%%%%%%%%%%%%%%%%%%%%%%%%%%%%%%%%%%%%%%%%%%%
\section{Introduction}
%%%%%%%%%%%%%%%%%%%%%%%%%%%%%%%%%%%%%%%%%%%%%%%%%%%%%%%%%%%%%%%%%
\label{sec:intro}

The Grand Unified Theories (GUTs) \cite{GUT} qualify among the most appealing physics scenarios beyond the Standard Model (SM) of electroweak and strong interactions. Though being under scrutiny for about 35 years they still attract a lot of attention across the high energy community due to their intrinsic predictivity
and to their potential for understanding the origin of our low energy world texture.
Apart from offering definite experimental motivations for e.g.~proton
decay or monopole searches, GUTs typically give rise to non-trivial
correlations among observables associated to different SM sectors.
The most prominent of these is the consistent determination of the
weak-mixing angle and the strong coupling arising from
the gauge coupling unification in a weak scale supersymmetric scenario.

In recent years, an extra boost to the field was triggered by the
discovery of non-zero neutrino masses in the sub-eV region. Within the
grand-unified scenarios this discovery translates into constraints
on the intermediate scales (typically well separated from the unification scale $M_G\sim 10^{16}$ GeV) underpinning some variant of the seesaw mechanism \cite{seesawI,seesawII}.
Furthermore, the observed peculiarity of
the lepton mixing pattern \cite{neutrinodata}
challenges the flavour
structure of the simplest models due to the strong correlations in the Yukawa sector.
In this respect, the requirement of minimality, that
stands for the simplicity of the relevant Higgs sector,
is a valuable guiding principle for model building.

On this basis, it has been argued recently that the minimal
supersymmetric $SO(10)$ model~\cite{MSSO10orig,MSSO10recent,Bajc:2004xe}
is indeed incompatible with the electroweak flavour constraints~\cite{MSSO10notworking}.
The minimal supersymmetric setting suffers from an inherent proximity of the GUT and the seesaw scales, at odds with the lower bound on the neutrino mass scale implied by the oscillation phenomena.
The proposed ways out (resorting e.g.~to a non-minimal Higgs sector \cite{MSGUT120} or invoking split supersymmetry~\cite{Bajc:2008dc}) hardly pair the appeal of the minimal setting.

Were a large (GUT scale)
breaking of global supersymmetry be at play (a possible
LHC test of this hypothesis has been recently put forward in Ref.~\cite{Hall:2009nd}), then baryon number violating
$d=5$ operators decouple from our low-energy world
and gauge unification exhibits naturally the
required splitting between the seesaw and the GUT scales~\cite{Gipson:1984aj,Deshpande:1992au,Bajc:2005zf,Bertolini:2009qj}. Nevertheless,
devising a realistic and simple enough $SO(10)$ GUT along these lines
remains a rather non-trivial task.

The main reason has to do with the structure of the minimal
Higgs sector of nonsupersymmetric $SO(10)$ models.
A full breaking of the GUT symmetry down to the SM can be achieved
via a pair of Higgs multiplets\footnote{The authors in Ref.~\cite{Babu:2005gx} observe that a one-step breaking
of $SO(10)$ can be achieved via one $144$ ($\overline{144}$) Higgs representation. Such a setting, suitable for a supersymmetric gauge unification, requires an extended matter
sector, including $45$ and $120$ multiplets, in order to accommodate realistic fermion masses~\cite{Nath:2009nf}.}:
one $45$-dimensional adjoint representation, $45_H$, and
one $16$-dimensional spinorial representation, $16_H$ (or one
126-dimensional tensor representation $126_H$).
A SM preserving breaking pattern is controlled by two $45_H$ vacuum expectation values (VEVs) and one $16_H$ (or $126_H$) VEV.
Different configurations of the two adjoint VEVs preserve
different $SO(10)$ subalgebras,
namely,
$4_{C}\, 2_{L}\, 1_{R}$,
(short-hand notation for
$SU(4)_c\otimes SU(2)_L\otimes U(1)_R$),
$ 3_{c}\, 2_{L}\, 2_{R}\, 1_{B-L}$,
$3_{c}\, 2_{L}\,1_{R}\,1_{B-L}$,
and the flipped or standard
$SU(5) \otimes U(1)$.
Except for the latter case, the subsequent breaking to the SM is obtained
via the standard $SU(5)$ conserving $16_H$ (or $126_H$) VEV.

Remarkably enough,  a consistent $SO(10)$ gauge symmetry breaking in the usual low-scale supersymmetric context requires minimally $45_H\oplus 54_H$ \cite{Aulakh:2000sn} (or $210_H$ \cite{MSSO10recent} in the renormalizable variant), in addition to
$\overline{16}_H \oplus 16_H$ (or $\overline{126}_H\oplus 126_H$).

The phenomenologically favored scenarios allowed by gauge
coupling unification
correspond minimally to a two-step breaking along one of the following directions~\cite{Bertolini:2009qj}:
\begin{eqnarray}
\label{chainVIII}
SO(10)&\stackrel{M_G}{\longrightarrow}
& 3_{c}\, 2_{L}\, 2_{R}\, 1_{B-L}\stackrel{M_I}{\longrightarrow} \mbox{SM}
%3_{c}\, 2_{L}\, 1_{R}\, 1_{B-L}
\,,\\[1ex]
\label{chainXII}
SO(10)&\stackrel{M_G}{\longrightarrow}
&4_{C}\, 2_{L}\, 1_{R}\stackrel{M_I}{\longrightarrow} \mbox{SM}
%3_{c}\, 2_{L}\, 1_{R}\, 1_{B-L}
\,,
\end{eqnarray}
where the first breaking stage is driven by the $45_H$ VEVs, while the breaking to the SM at the intermediate scale $M_I$, several orders of magnitude below the unification scale $M_G$, is controlled
by the $16_H$ (or $126_H$) VEV. One of the two $45_H$ VEVs may also contribute to the second step (see the discussion on the
required intermediate scale Higgs multiplets in Ref.~\cite{Bertolini:2009qj} and in \sect{subsec:ESH}).

Gauge unification, even without proton decay limits, excludes any
intermediate $SU(5)$-symmetric stages. On the other hand,
a series of studies in the early 1980's of the $45_H\oplus 16_H$ model \cite{Yasue,Anastaze:1983zk,Babu:1984mz} indicated that
the only intermediate stages allowed by the scalar sector dynamics were
the flipped $SU(5)\otimes U(1)$ for leading $45_H$ VEVs or the standard $SU(5)$ GUT
for  dominant $16_H$ VEV.

This observation excluded
the simplest $SO(10)$ Higgs sector from realistic consideration.

In this paper we show that the exclusion of the breaking patterns
in \eqs{chainVIII}{chainXII} is an artifact of the tree level potential. As a matter of fact,
some entries of the scalar hessian
are accidentally
over-constrained at the tree level. A number of scalar interactions that, by a simple inspection of the relevant global symmetries and their explicit breaking, are expected to contribute to these critical entries, are not effective at the tree level.

On the other hand, once quantum corrections are considered,
contributions of $O(M_G^2/16\pi^2)$ induced on these entries
open in a natural way
all group-theoretically allowed vacuum configurations.
Remarkably enough, the study of the one-loop effective potential
can be consistently carried out just for the critical tree level hessian entries (that correspond to specific pseudo-Goldstone boson masses).
For all other states in the scalar spectrum, quantum corrections
remain perturbations of the tree level results and do not affect
the discussion of the vacuum pattern.

Our conclusions apply to any Higgs setting where the first step
of the $SO(10)$ gauge symmetry breaking is driven by the $45_H$ VEVs,
while the other Higgs representations control the
intermediate and weak scale stages.
The results presented here and in Ref.~~\cite{Bertolini:2009qj}
do open the path towards a realistic
nonsupersymmetric $SO(10)$ unification. A detailed study of minimal setups will be the subject of a future work.

The paper is organized as follows.
The study of the tree-level scalar potential and the related scalar mass spectrum are concisely reviewed in \sects{sect:minimalSO10}{sect:classicalvacuum}.
A detailed understanding of the mass textures is developed in \sect{sect:understanding} in terms of a systematic discussion of the accidental global symmetries and the associated pseudo-Goldstone bosons.
In \sect{sec:quantumvacuum} we calculate the relevant quantum corrections
by means of the one-loop effective potential, and we prove the
existence of the new vacua.
% for natural values of the parameters.
The main results and the prospects for further developments are summarized in \sect{sec:outlook}.
Most of the technical aspects of the work are deferred to \apps{app:so10algebra}{app:1Lmasses}.

%%%%%%%%%%%%%%%%%%%%%%%%%%%%%%%%%%%%%%%%%%%%%%%%%%%%%%%%%%%%%%%%
\section{The minimal SO(10) GUT}
%%%%%%%%%%%%%%%%%%%%%%%%%%%%%%%%%%%%%%%%%%%%%%%%%%%%%%%%%%%%%%%%
\label{sect:minimalSO10}

In this study we consider a nonsupersymmetric $SO(10)$ setup
featuring the minimal Higgs content sufficient to trigger the spontaneous breakdown of the GUT symmetry down to the
standard electroweak model.
Minimally, the scalar sector spans over a reducible $45_{H}\oplus 16_{H}\oplus 10_{H}$ representation. The adjoint $45_{H}$ and the spinor $16_{H}$ multiplets contain three SM singlets that may acquire GUT
scale VEVs.

The $10_{H}$, which together with the relevant components of $16_{H}$ triggers the $\mbox{SM} \to SU(3)_{c}\otimes U(1)_{Q}$ breaking,
is introduced in order to admit for a potentially realistic fermionic spectrum.
The VEV of $10_{H}$ is very tiny in comparison with the VEVs of $45_{H}$ or $16_{H}$ and in the chains we are considering it mixes with $16_{H}$ only at the electroweak scale. On the other hand, most of the $10_{H}$ component fields do maintain a natural mass of the order of the unification scale. In this respect they play a role also for the details of the theory at the GUT scale.
Nevertheless, as we shall see, the $10_H$ is not needed for the scope of the present discussion and we shall neglect it altogether.

Let us emphasize once more that the issue we shall be dealing with is inherent to all nonsupersymmetric $SO(10)$ models with one adjoint $45_H$ governing the first breaking step. Only one additional scalar representation interacting with the adjoint is sufficient to demonstrate conclusively our claim.
In this respect, the choice of the $SO(10)$ spinor to trigger the intermediate symmetry breakdown is a mere convenience and a similar line of reasoning can be devised for the scenarios in which $B-L$ is broken for instance by a 126-dimensional $SO(10)$ tensor.

We shall therefore study the structure of the vacua of a $SO(10)$ Higgs potential with only the $45_{H} \oplus 16_{H}$ representation at play.
Following the common convention, we define $16_{H} \equiv \chi$ and denote by $\chi_{+}$ and $\chi_{-}$ the multiplets transforming as positive and negative chirality components of the reducible 32-dimensional $SO(10)$ spinor representation respectively
Similarly, we shall use the symbol $\Phi$ (or the derived $\phi$, c.f. \app{app:so10algebra})
for the adjoint Higgs representation $45_{H}$ (or its components in the natural basis).
%We refer the reader to \app{app:so10algebra} for details.

The minimal $SO(10)$ GUT accommodates the SM matter in three copies of $SO(10)$ spinors $16_{F}^{i}$, ($i=1,2,3$).
%In analogy to the $10_{H}$ Higgs representation,
The fermions (and their Yukawa interactions) do not play any role in the GUT scale dynamics and will not be considered further (we assume the masses of the right-handed neutrinos to be small with respect to the unification scale).
The detailed study of realistic Higgs and Yukawa sectors will be the
subject of a forthcoming paper.

%=====================================================================
\subsection{The tree-level Higgs potential}
%=====================================================================
The most general renormalizable tree-level scalar potential which
can be constructed out of $45_H$ and $16_H$ reads (see for instance
Refs.~\cite{Li:1973mq,Buccella:1980qb}):
\beq
V_0=V_{\Phi}+V_{\chi}+V_{\Phi\chi} \, ,
\label{potentialV0}
\eeq
where, according to the notation in \app{app:so10algebra},
\bea
\label{potentialV45}
V_{\Phi}&=&
-\frac{\mu^2}{2}\Tr\Phi^2 + \frac{a_1}{4}(\Tr\Phi^2)^2 + \frac{a_2}{4}\Tr\Phi^4 \, , \\
\label{potentialV16}
V_{\chi}&=&
-\frac{\nu^2}{2}\chi^\dag\chi
+\frac{\lambda_1}{4}(\chi^\dag\chi)^2
 +\frac{\lambda_2}{4}(\chi_+^\dag\Gamma_j\chi_-)(\chi_-^\dag\Gamma_j\chi_+) \,\nn
 \eea
and
\beq
\label{potentialV4516}
V_{\Phi\chi}=
\alpha(\chi^\dag\chi)\Tr\Phi^2+\beta\chi^\dag\Phi^2\chi
+\tau\chi^\dag\Phi\chi \, .
\eeq
The mass terms and coupling constants above are real by hermiticity.
Linear and cubic $\Phi$ self-interactions are absent
due the zero trace of the $SO(10)$ adjoint representation.
For the sake of simplicity, all tensorial indices have been suppressed.
%The interested reader may find more information on the tensorial structure of $V_{0}$ in \app{app:so10algebra}.

%=========================================================================
\subsection{The symmetry breaking patterns}
\label{sec:breakingpatterns}
%=========================================================================

%-------------------------------------------------------------------------
\subsubsection{The SM singlets}
\label{sec:SMsinglet}
%-------------------------------------------------------------------------

There are in general three SM singlets in the $45_H\oplus16_H$ representation of $SO(10)$. Using $(B-L)/2\equiv X$
and labeling the field components according to
$3_{c}\, 2_{L}\, 2_{R}\, 1_X$, the SM singlets
reside in the $(1,1,1,0)$ and $(1,1,3,0)$ submultiplets of $45_{H}$
and in the $(1,1,2,+\tfrac{1}{2})$ component of $16_{H}$.
We denote their VEVs as
\begin{align}
\label{vevs}
&\vev{(1,1,1,0)}\equiv \omega_{Y}, \nn \\[0.5ex]
&\vev{(1,1,3,0)}\equiv \omega_{R},  \\[0.5ex]
&\vev{(1,1,2,+\tfrac{1}{2})}\equiv \chi_{R}, \nn
\end{align}
where $\omega_{Y,R}$ are real and $\chi_{R}$ can be taken real by
a phase redefinition of the $16_H$.
Different VEV configurations
trigger the spontaneous breakdown of the $SO(10)$ symmetry into a number of subgroups. Namely, for $\chi_{R}= 0$ one finds
\begin{align}
\label{vacua}
&\omega_{R}= 0,\, \omega_{Y}\neq 0\; : & 3_c\, 2_L\, 2_R\, 1_X \nn \\[0.5ex]
&\omega_{R}\neq 0,\, \omega_{Y}= 0\; : & 4_{C} 2_L 1_R \nn \\[0.5ex]
&\omega_{R}\neq 0,\, \omega_{Y}\neq 0\; : & 3_c\, 2_L\, 1_R\, 1_X   \\[0.5ex]
&\omega_{R}=-\omega_{Y}\neq 0\; : & \mbox{flipped}\, 5'\, 1_{Z'} \nn \\[0.5ex]
&\omega_{R}=\omega_{Y}\neq 0\; :  & \mbox{standard}\, 5\, 1_{Z} \nn
\end{align}
with  $5\, 1_{Z}$ and $5'\, 1_{Z'}$ standing for the two different
embedding of the $SU(5)$ subgroup into $SO(10)$, i.e. standard and ``flipped'' respectively (see the discussion at the end of the section).

When $\chi_{R}\neq 0$ all intermediate gauge symmetries are spontaneously broken down to the SM group, with the exception of the last case which maintains the standard $SU(5)$ subgroup unbroken and will no further be considered.

The classification in \eq{vacua} depends on the phase conventions used in the parametrization of the SM singlet subspace of $45_{H} \oplus 16_{H}$.
The statement that $\omega_{R}=\omega_{Y}$ yields the standard $SU(5)$
vacuum while  $\omega_{R}=-\omega_{Y}$ corresponds to the flipped setting defines a particular basis in this subspace
(see \sect{sec:su5vsflippedsu5}).
The consistency of any chosen framework is then verified against the corresponding Goldstone boson spectrum.

The decomposition of the $45_H$ and $16_H$ representations with respect to the relevant $SO(10)$ subgroups is detailed in Tables \ref{tab:16decomp} and \ref{tab:45decomp}.

\renewcommand{\arraystretch}{1.3}
\begin{table*}
\begin{tabular}{lllllccc}%{llllllllr}
\hline \hline
%$10$ %$SO(10)$
 $4_C\,2_L\,2_R $
& $4_C\,2_L\,1_R $
& $3_c\,2_L\,2_R\,1_{X} $
& $3_c\,2_L\,1_R\,1_{X} $
& $3_c\,2_L\,1_Y $
& $5$ %$SU(5)$
& $5'\,1_{Z'}$ %$SU(5)'\otimes U(1)_{Z} $
& $1_{Y'}$ %$U(1)_{Y'}$
\\
\hline
%$16$
$\left({4,2,1} \right)$
& $\left({4,2,0} \right)$
& $\left({ 3,2,1},+\frac{1}{6} \right)$
& $\left({ 3,2},0,+\frac{1}{6} \right)$
& $\left({ 3,2},+\frac{1}{6} \right)$
& $\left.{ 10}\right.$
& $\left({ 10},+1 \right)$
& $\left. +\frac{1}{6} \right.$
\\
\null
&
& $\left({ 1,2,1},-\frac{1}{2} \right)$
& $\left({ 1,2},0,-\frac{1}{2} \right)$
& $\left({ 1,2},-\frac{1}{2} \right)$
& $\left.{ \overline{5}} \right.$
& $\left({ \overline{5}},-3 \right)$
& $\left. -\frac{1}{2} \right.$
\\
$\left({ \overline{4},1,2} \right)$
& $\left({ \overline{4},1,+\frac{1}{2}} \right)$
& $\left({ \overline{3},1,2},-\frac{1}{6} \right)$
& $\left({ \overline{3},1},+\frac{1}{2},-\frac{1}{6} \right)$
& $\left({ \overline{3},1},+\frac{1}{3} \right)$
& $\left.{ \overline{5}} \right.$
& $\left({ 10},+1 \right)$
& $\left. -\frac{2}{3} \right.$
\\
\null
& $\left({ \overline{4},1,-\frac{1}{2}} \right)$
&
& $\left({ \overline{3},1},-\frac{1}{2},-\frac{1}{6} \right)$
& $\left({ \overline{3},1},-\frac{2}{3} \right)$
& $\left.{ 10} \right.$
& $\left({ \overline{5}},-3 \right)$
& $\left.  +\frac{1}{3} \right.$
\\
\null
&
& $\left({ 1,1,2},+\frac{1}{2} \right)$
& $\left({ 1,1},+\frac{1}{2},+\frac{1}{2} \right)$
& $\left({ 1,1},+1 \right)$
& $\left.{ 10} \right.$
& $\left({ 1},+5 \right)$
& $\left. 0 \right.$
\\
\null
&
&
& $\left({ 1,1},-\frac{1}{2},+\frac{1}{2} \right)$
& $\left({ 1,1},0 \right)$
& $\left.{ 1} \right.$
& $\left({ 10},+1 \right)$
& $\left. +1 \right.$
\\
\hline \hline
\end{tabular}
\mycaption{Decomposition of the spinorial representation  $16$ with respect to the various  $SO(10)$ subgroups. The definitions and normalization of the abelian charges are given in the text.}
\label{tab:16decomp}
\end{table*}

\renewcommand{\arraystretch}{1.3}
\begin{table*}
\begin{tabular}{lllllccc}%{llllllllr}
\hline \hline
%$10$ %$SO(10)$
 $4_C\,2_L\,2_R $
& $4_C\,2_L\,1_R $
& $3_c\,2_L\,2_R\,1_{X} $
& $3_c\,2_L\,1_R\,1_{X} $
& $3_c\,2_L\,1_Y $
& $5$ %$SU(5)$
& $5'\,1_{Z'}$ %$SU(5)'\otimes U(1)_{Z} $
& $1_{Y'}$ %$U(1)_{Y'}$
\\
\hline
 $\left({ 1,1,3} \right)$
& $\left({ 1,1},+1 \right)$
& $\left({ 1,1,3},0 \right)$
& $\left({ 1,1},+1,0 \right)$
& $\left({ 1,1},+1 \right)$
& $\left.{ 10} \right.$
& $\left({ 10},-4 \right)$
& $\left. +1 \right.$
\\
\null
& $\left({ 1,1},0 \right)$
&
& $\left({ 1,1},0,0 \right)$
& $\left({ 1,1},0 \right)$
& $\left.{ 1} \right.$
& $\left({ 1},0 \right)$
& $\left. 0 \right.$
\\
\null
& $\left({ 1,1},-1 \right)$
&
& $\left({ 1,1},-1,0 \right)$
& $\left({ 1,1},-1 \right)$
& $\left.{ \overline{10}} \right.$
& $\left({ \overline{10}},+4 \right)$
& $\left. -1 \right.$
\\
$\left({ 1,3,1} \right)$
& $\left({ 1,3},0 \right)$
& $\left({ 1,3,1},0 \right)$
& $\left({ 1,3},0,0 \right)$
& $\left({ 1,3},0 \right)$
& $\left.{ 24} \right.$
& $\left({ 24},0 \right)$
& $\left. 0 \right.$
\\
$\left({ 6,2,2} \right)$
& $\left({ 6,2},+\frac{1}{2} \right)$
& $\left({ 3,2,2},-\frac{1}{3} \right)$
& $\left({ 3,2},+\frac{1}{2},-\frac{1}{3} \right)$
& $\left({ 3,2},\frac{1}{6} \right)$
& $\left.{ 10} \right.$
& $\left({ 24},0 \right)$
& $\left. -\frac{5}{6} \right.$
\\
\null
& $\left({ 6,2},-\frac{1}{2} \right)$
&
& $\left({ 3,2},-\frac{1}{2},-\frac{1}{3} \right)$
& $\left({ 3,2},-\frac{5}{6} \right)$
& $\left.{ 24} \right.$
& $\left({ 10},-4 \right)$
& $\left. +\frac{1}{6} \right.$
\\
\null
&
& $\left({ \overline{3},2,2},+\frac{1}{3} \right)$
& $\left({ \overline{3},2},+\frac{1}{2},+\frac{1}{3} \right)$
& $\left({ \overline{3},2},+\frac{5}{6} \right)$
& $\left.{ 24} \right.$
& $\left({ \overline{10}},+4 \right)$
& $\left. -\frac{1}{6} \right.$
\\
\null
&
&
& $\left({ \overline{3},2},-\frac{1}{2},+\frac{1}{3} \right)$
& $\left({ \overline{3},2},-\frac{1}{6} \right)$
& $\left.{ \overline{10}} \right.$
& $\left({ 24},0 \right)$
& $\left. +\frac{5}{6} \right.$
\\
$\left({ 15,1,1} \right)$
& $\left({ 15,1},0 \right)$
& $\left({ 1,1,1},0 \right)$
& $\left({ 1,1},0,0 \right)$
& $\left({ 1,1},0 \right)$
& $\left.{ 24} \right.$
& $\left({ 24},0 \right)$
& $\left. 0 \right.$
\\
\null
&
& $\left({ 3,1,1},+\frac{2}{3} \right)$
& $\left({ 3,1},0,+\frac{2}{3} \right)$
& $\left({ 3,1},+\frac{2}{3} \right)$
& $\left.{ \overline{10}} \right.$
& $\left({ \overline{10}},+4 \right)$
& $\left. +\frac{2}{3} \right.$
\\
\null
&
& $\left({ \overline{3},1,1},-\frac{2}{3} \right)$
& $\left({ \overline{3},1},0,-\frac{2}{3} \right)$
& $\left({ \overline{3},1},-\frac{2}{3} \right)$
& $\left.{ 10} \right.$
& $\left({ 10},-4 \right)$
& $\left. -\frac{2}{3} \right.$
\\
\null
&
& $\left({ 8,1,1},0 \right)$
& $\left({ 8,1},0,0 \right)$
& $\left({ 8,1},0 \right)$
& $\left.{ 24} \right.$
& $\left({ 24},0 \right)$
& $\left. 0 \right.$
\\
\hline \hline
\end{tabular}
\mycaption{Same as in Table \ref{tab:16decomp} for the $SO(10)$ adjoint ($45$) representation.}
\label{tab:45decomp}
\end{table*}

%-------------------------------------------------------------------------
\subsubsection{The L-R chains}
\label{sec:L-Rchains}
%-------------------------------------------------------------------------

According to the analysis in Ref.~\cite{Bertolini:2009qj},
the potentially viable breaking chains fulfilling the basic gauge
unification constraints (with a minimal $SO(10)$ Higgs sector)
correspond to the settings:
\begin{multline}
\omega_{Y}\gg \omega_{R} > \chi_{R}\ :  \\[1ex]
SO(10)\to 3_{c}2_{L}2_{R}1_{X}\to 3_{c}2_{L}1_{R}1_{X}\to 3_{c}2_{L}1_{Y}
\end{multline}
\begin{multline}
\omega_{R}\gg \omega_{Y}\ > \chi_{R}\ :   \\[1ex]
SO(10)\to 4_{C}2_{L}1_{R}\to 3_{c}2_{L}1_{R}1_{X}\to 3_{c}2_{L}1_{Y}
\end{multline}

As remarked in \cite{Bertolini:2009qj}, the cases
$\chi_{R} \sim \omega_{R}$ or $\chi_{R} \sim \omega_{Y}$ lead to effective two-step $SO(10)$ breaking patterns with a non-minimal
set of surviving scalars at the intermediate scale.
On the other hand, a truly two-step setup can be recovered (with a minimal fine tuning) by considering the cases where
$\omega_{R}$ or $\omega_{Y}$ exactly vanish. Only the explicit study
of the scalar potential determines which of the textures are allowed.

We have verified that in all cases the GUT thresholds effects related to the relevant pseudo-Goldstone mass patterns obtained in the present analysis fully comply with the unification constraints in Ref.~\cite{Bertolini:2009qj}. Furthermore, the lower bounds on the position of the $B-L$ scale are consistently increased, hence improving the prospects for a successful model building.

%-------------------------------------------------------------------------
\subsubsection{Standard SU(5) versus flipped SU(5)}
\label{sec:su5vsflippedsu5}
%-------------------------------------------------------------------------

There are in general two distinct SM-compatible embeddings  of  $SU(5)$
into $SO(10)$~\cite{DeRujula:1980qc,Barr:1981qv}.
They differ in one generator of the $SU(5)$ Cartan algebra and
therefore in the $U(1)_Z$ cofactor.

In the ``standard'' embedding, the weak hypercharge operator
$
Y=T^{(3)}_R+T_X
$
belongs to the $SU(5)$ algebra and the orthogonal Cartan generator
$Z$ (obeying $ [T_{i},Z]=0$ for all $T_{i}\in SU(5)$) is given by
$
Z =-4T^{(3)}_R+6T_X
$.

In the ``flipped'' $SU(5)'$ case, the right-handed isospin
assignment of quark and leptons into the $SU(5)'$ multiplets
is turned over so that the ``flipped'' hypercharge generator reads
$Y'=-T^{(3)}_R+T_X$.
Accordingly, the additional $U(1)_{Z'}$ generator reads
$Z' =4T^{(3)}_R+6T_X$,
such that $ [T_{i},Z']=0$ for all $T_{i}\in SU(5)'$.
Weak hypercharge is then given by
$Y=(Z'-Y')/5$.

\Tables{tab:16decomp}{tab:45decomp} show
the standard and flipped $SU(5)$ decompositions of the spinorial and adjoint $SO(10)$ representations respectively.

The two $SU(5)$ vacua in \eq{vacua} differ
by the texture of the adjoint representation VEVs:
in the standard $SU(5)$ case they are aligned with the $Z$ operator
while they match the $Z'$ structure
in the flipped $SU(5)'$ setting (see \app{app:explicitgenerators} for
an explicit representation).
\vskip 1mm

%%%%%%%%%%%%%%%%%%%%%%%%%%%%%%%%%%%%%%%%%%%%%%%%%%%%%%%%%%%%%%%%
\section{The classical vacuum}
%%%%%%%%%%%%%%%%%%%%%%%%%%%%%%%%%%%%%%%%%%%%%%%%%%%%%%%%%%%%%%%%
\label{sect:classicalvacuum}

%=========================================================================
\subsection{The stationarity conditions}
%=========================================================================

By substituting \eq{vevs} into \eq{potentialV0} the vacuum manifold reads
\begin{align}
\vev{V_0}=&-2\mu^2(2\omega_R^2 + 3\omega_Y^2) + 4a_1(2\omega_R^2 +3\omega_Y^2)^2 \nn \\[1ex]
&+\frac{a_2}{4}(8\omega_R^4 + 21\omega_Y^4 + 36\omega_R^2\omega_Y^2) \nn \\[1ex]
&-\frac{\nu^2}{2}\chi_R^2 + \frac{\lambda_1}{4}\chi_R^4  \nn \\[1ex]
&+4\alpha\chi_R^2(2\omega_R^2 +3\omega_Y^2)
+\frac{\beta}{4}\chi_R^2(2\omega_R + 3\omega_Y)^2 \nn \\[1ex]
&-\frac{\tau}{2}\chi_R^2(2\omega_R + 3\omega_Y)
\end{align}

The corresponding three stationary conditions can be conveniently written as

\bwt
\begin{align}
& \frac{1}{8}\left(\frac{\partial \vev{V_0}}{\partial\omega_R}
-\frac{2}{3}\frac{\partial \vev{V_0}}{\partial\omega_Y}\right)=
\left[ -\mu^2+4a_1(2\omega_R^2 + 3\omega_Y^2)
%\nn \\[1ex]
+ \frac{a_2}{4}
(4\omega_R^2+7\omega_Y^2-2\omega_Y\omega_R)+2\alpha\chi_R^2\right] (\omega_R-\omega_Y) = 0
\label{eqstatmu} \\[1ex]
& \omega_Y\frac{\partial \vev{V_0}}{\partial\omega_R}
-\omega_R \frac{2}{3}\frac{\partial \vev{V_0}}{\partial\omega_Y} =
-\left[ 4a_2(\omega_R+\omega_Y)\omega_R\omega_Y \right.
%\nn \\[1ex]
+ \left. \beta\chi_R^2(2\omega_R + 3\omega_Y)
-\tau\chi_R^2\right](\omega_R-\omega_Y) = 0
\label{eqstat0} \\[1ex]
& \frac{\partial \vev{V_0}}{\partial\chi_R}=
\left[ -\nu^2+\lambda_1\chi_R^2+8\alpha(2\omega_R^2 + 3\omega_Y^2)
%\nn \\[0.5ex]
+ \frac{\beta}{2}(2\omega_R + 3\omega_Y)^2-\tau(2\omega_R + 3\omega_Y) \right] \chi_R = 0
\label{eqstatnu}
\end{align}
\ewt

We have chosen linear combinations that factor out the
uninteresting standard $SU(5)\otimes U(1)_Z$ solution, namely $\omega_R=\omega_Y$.

In summary, when $\chi_R = 0$, \eqs{eqstatmu}{eqstat0} allow for four possible vacua:
\begin{itemize}
\item{$\omega = \omega_R = \omega_Y$ (standard $5\, 1_{Z}$)}
\item{$\omega = \omega_R = -\omega_Y$ (flipped $5'\, 1_{Z'}$)}
\item{$\omega_R=0$ and $\omega_Y \neq 0$ ($3_c\, 2_L\, 2_R\, 1_{X}$)}
\item{$\omega_R \neq 0$ and $\omega_Y = 0$ ($4_C\, 2_L\, 1_R$)}
\end{itemize}

As we shall see, the last two options are not {\em tree level} minima.
Let us remark that for $\chi_R \neq 0$,
%and $|\omega_Y/\omega_R|\neq 1$,
\eq{eqstat0} implies naturally a correlation
among the $45_H$ and $16_H$ VEVs,
%namely $\omega_Y \omega_R \sim \chi_R^2$,
or a fine tuned relation between $\beta$ and $\tau$, depending on the stationary solution.
In the cases $\omega_R=-\omega_Y$, $\omega_R=0$ and $\omega_Y = 0$
one obtains $\tau = \beta\omega$, $\tau = 3\beta\omega_Y$ and $\tau = 2\beta\omega_R$ respectively. Consistency with the scalar mass spectrum
must be verified in each case.

%=========================================================================
\subsection{The tree-level spectrum}
%=========================================================================

The gauge and scalar spectra corresponding to the SM vacuum configuration
(with non-vanishing VEVs in $45_{H}\oplus 16_{H}$) are detailed
in \app{app:Treemasses}.
%\app{app:gaugespectrum} and \ref{app:scalarspectrum}, respectively.

The scalar spectra obtained in various limits of the tree-level Higgs potential, corresponding to the appearance of accidental global symmetries, are derived in \apps{45only}{4516trivialinteraction}.
The emblematic case $\chi_R=0$ is scrutinized in \app{4516spectrumchi0}.

%=========================================================================
\subsection{Constraints on the potential parameters}
%=========================================================================
\label{furtherconstraints}

The parameters (couplings and VEVs) of the scalar potential are constrained by the requirements
of boundedness and the absence of tachyonic states, ensuring that
the vacuum is stable and the stationary points correspond to physical minima.

Necessary conditions for vacuum stability are derived
in \app{boundedpot}. In particular, on the $\chi_R=0$ section
one obtains
\begin{equation}
\label{strongestboundednessconstraints}
a_1 > -\tfrac{13}{80} a_2
%\, , \quad \lambda_1 > 0
\, .
\end{equation}

Considering the general case, the absence of tachyons in the scalar spectrum yields among else
\beq
\label{bound130810}
a_2 < 0 \, , \qquad -2<\omega_Y/\omega_R<-\tfrac{1}{2} \,.
\eeq
The strict constraint on $\omega_Y/\omega_R$
is a consequence of the tightly correlated form of the tree-level
masses of the $(8,1,0)$ and $(1,3,0)$ submultiplets of $45_{H}$,
labeled according to the SM ($3_c\, 2_L\, 1_Y$) quantum numbers, namely
\bea
\label{310PGBmass}
M^2(1,3,0) & = &
2 a_2 (\omega _Y - \omega _R) (\omega _Y + 2 \omega _R) \, , \\[1ex]
\label{810PGBmass}
M^2(8,1,0) & = &
2 a_2 (\omega _R - \omega _Y) (\omega _R + 2 \omega _Y) \,,
\eea
that are simultaneously positive only if \eq{bound130810} is enforced.
For comparison with previous studies, let us remark that in the $\tau=0$ limit (corresponding to an extra $Z_2$ symmetry
$\Phi\to -\Phi$) the intersection of the constraints from \eq{eqstat0}, \eqs{310PGBmass}{810PGBmass} and
the mass eigenvalues of the
$(1,1,1)$ and
$(3,2,1/6)$ states, yields
\beq
\label{boundYasue}
a_2 < 0 \, ,
%\quad \beta > 0\, ,
\quad  -1 \leq \omega_Y/\omega_R \leq -\tfrac{2}{3}
% \quad \text{and} \quad \beta>0
\, ,
\eeq
thus recovering the results of
Refs.~\cite{Yasue,Anastaze:1983zk,Babu:1984mz}.

In either case, one concludes by inspecting the scalar mass spectrum
that flipped $SU(5)'\otimes U(1)_{Z'}$ is for $\chi_R=0$
the only solution admitted by \eq{eqstat0} consistent with the
constraints in \eq{bound130810} (or \eq{boundYasue}).
For $\chi_R\neq 0$,
the fine tuned possibility of having or $\omega_Y/\omega_R \sim -1$ such that $\chi_R$ is obtained
at an intermediate scale fails to reproduce the SM couplings~\cite{Bertolini:2009qj}.
Analogous and obvious conclusions hold for
$\omega_Y \sim \omega_R \sim \chi_R \sim M_G$ and for
$\chi_R\gg \omega_{R,Y}$
 (standard $SU(5)$ in the first stage).

This is the origin of the common knowledge that nonsupersymmetric $SO(10)$ settings with the adjoint VEVs driving the gauge symmetry breaking are not phenomenologically viable. In particular, a large
hierarchy between the $45_H$ VEVs, that would set the stage for
consistent unification patterns, is excluded.

The key question is: why are the masses of the states in \eqs{310PGBmass}{810PGBmass} so tightly correlated?
Equivalently, why do they depend on $a_2$ only?

%%%%%%%%%%%%%%%%%%%%%%%%%%%%%%%%%%%%%%%%%%%%%%%%%%%%%%%%%%%%%%%%%%%%%%%%
\section{Understanding the scalar spectrum}
%%%%%%%%%%%%%%%%%%%%%%%%%%%%%%%%%%%%%%%%%%%%%%%%%%%%%%%%%%%%%%%%%%%%%%%%
\label{sect:understanding}

A detailed comprehension of the patterns in the scalar spectrum
may be achieved by understanding the correlations between mass textures and the symmetries of the scalar potential. In particular, the
appearance of accidental global symmetries in limiting cases
may provide the rationale for the dependence of mass eigenvalues from specific couplings.
To this end we classify the most interesting cases, providing
a counting of the would-be Goldstone bosons (WGB)
and pseudo Goldstone bosons (PGB) for each case.
A side benefit of this discussion is a consistency check of the
explicit form of the mass spectra.

%-------------------------------------------------
\subsection{45 only with $a_2=0$}
%-------------------------------------------------
\label{45a2}

Let us first consider the potential generated by $45_H$, namely
$V_\Phi$ in \eq{potentialV0}.
When $a_2=0$,
i.e. when only trivial $45_H$ invariants (built off moduli) are considered, the scalar potential
exhibits an enhanced global symmetry: $O(45)$.
The spontaneous symmetry breaking (SSB) triggered by the $45_H$ VEV
reduces the global symmetry to $O(44)$.
As a consequence, 44 massless states are expected in the scalar spectrum.
This is verified explicitly in \app{45only}.
Considering the case of the $SO(10)$ gauge symmetry broken to the flipped $SU(5)'\otimes U(1)_{Z'}$, $45-25=20$ WGB, with the quantum numbers
of the coset $SO(10)/SU(5)'\otimes U(1)_{Z'}$ algebra,
decouple from the physical spectrum
while, $44-20=24$ PGB remain, whose
mass depends on the explicit breaking term $a_2$.

%-------------------------------------------------
\subsection{16 only with $\lambda_2=0$}
%-------------------------------------------------
\label{16lambda2}

We proceed in analogy with the previous discussion.
Taking $\lambda_{2}=0$ in $V_\chi$
enhances the global symmetry to $O(32)$.
The spontaneous breaking of $O(32)$ to $O(31)$
due to the $16_H$  VEV leads to
31 massless modes, as it is explicitly seen in \app{16only}.
Since the gauge $SO(10)$ symmetry is broken by $\chi_R$ to the standard $SU(5)$, $45-24=21$ WGB, with the quantum numbers
of the coset $SO(10)/SU(5)$ algebra,
decouple from the physical spectrum,
while $31-21=10$ PGB do remain.
Their masses depend on the explicit
breaking term $\lambda_2$.

%-------------------------------------------------
\subsection{A trivial 45-16 potential $(a_2=\lambda_2=\beta=\tau=0)$}
%-------------------------------------------------
\label{therelevantone}

When only trivial invariants (i.e. moduli) of both $45_H$ and $16_H$ are considered,
the global symmetry of $V_{0}$ in \eq{potentialV0}
is $O(45)\otimes O(32)$.
This symmetry is spontaneously broken into $O(44)\otimes O(31)$ by the $45_H$ and $16_H$ VEVs
yielding 44+31=75 GB in the scalar spectrum (see \app{4516justnorms}).
Since in this case, the gauge $SO(10)$ symmetry is broken to the SM gauge group,
$45-12=33$ WGB, with the quantum numbers
of the coset $SO(10)/SM$ algebra,
decouple from the physical spectrum,
while $75-33=42$ PGB remain.
Their masses are generally expected to receive contributions
from the explicitly breaking terms $a_2$, $\lambda_2$, $\beta$ and $\tau$.

%-------------------------------------------------
\subsection{A trivial 45-16 interaction $(\beta=\tau=0)$}
%-------------------------------------------------
\label{45-16betatau}

Turning off just the $\beta$ and $\tau$ couplings still allows for
independent global rotations of the $\Phi$ and $\chi$ Higgs fields.
The largest global symmetries are those determined
by the $a_2$ and $\lambda_2$ terms
in $V_{0}$, namely $O(10)_{45}$ and $ O(10)_{16}$, respectively.
Consider the spontaneous breaking to global flipped
$SU(5)'\otimes U(1)_{Z'}$ and the standard $SU(5)$
by the $45_H$ and $16_H$ VEVs, respectively.
This setting gives rise to  $20 + 21 = 41$ massless scalar modes.
The gauged $SO(10)$ symmetry is broken
to the SM group so that 33 WGB decouple from the physical spectrum.
Therefore, 41-33=8 PGB remain,
whose masses receive contributions from the explicit
breaking terms $\beta$ and $\tau$.
All of these features are readily verified by inspection of the scalar mass spectrum in \app{4516trivialinteraction}.

%=========================================================================
\subsection{A tree-level accident}
%=========================================================================

The tree-level masses of the crucial $(1,3,0)$ and $(8,1,0)$ multiplets belonging to the $45_H$
depend only on the parameter $a_2$ but {\em not} on the other parameters expected (c.f. \ref{therelevantone}),
namely $\lambda_2$, $\beta$ and $\tau$.

While the $\lambda_2$ and $\tau$ terms cannot obviously contribute
at the tree level to $45_H$ mass terms,
one would generally expect a contribution from the $\beta$ term, proportional to $\chi_R^2$.
Using the parametrization $\Phi=\sigma_{ij}\phi_{ij}/4$, where the $\sigma_{ij}$  ($i,j\in\{1,..,10\}$, $i\neq j$) matrices
represent the $SO(10)$ algebra on the 16-dimensional spinor basis
(c.f. \app{app:so10algebra}), one obtains a $45_H$ mass term of the form
\beq
\label{45massbeta}
\frac{\beta}{16}\chi_R^2\ (\sigma_{ij})_{16\beta}(\sigma_{kl})_{\beta 16}\ \phi_{ij}\phi_{kl} \, .
\eeq
The projection of the $\phi_{ij}$ fields onto the
$(1,3,0)$ and $(8,1,0)$ components lead, as we know, to vanishing
contributions.

This result can actually be understood on general grounds
by observing that
the scalar interaction in \eq{45massbeta} has the same structure as the
gauge boson mass from the covariant-derivative interaction with the $16_H$, c.f. \eq{fielddepmass16}.
As a consequence, no tree-level mass contribution from the $\beta$ coupling can be generated  for the $45_H$
scalars carrying the quantum numbers of the standard $SU(5)$ algebra.

This behavior can be again verified by inspecting the relevant scalar spectra in \app{app:scalarspectrum}.

The above considerations provide a clear rationale for the
accidental tree level constraint on $\omega_Y/\omega_R$,
that holds independently on the size of $\chi_R$.

On the other hand, we should expect the $\beta$ and $\tau$ interactions to contribute $O(M_G/4\pi)$ terms to the masses of $(1,3,0)$ and $(8,1,0)$ at the quantum level.

Similar contributions should also arise from the gauge interactions,
that break explicitly the independent global transformations on
the $45_H$ and $16_H$ discussed in the previous subsections.

\begin{figure}
\includegraphics[width=5cm]{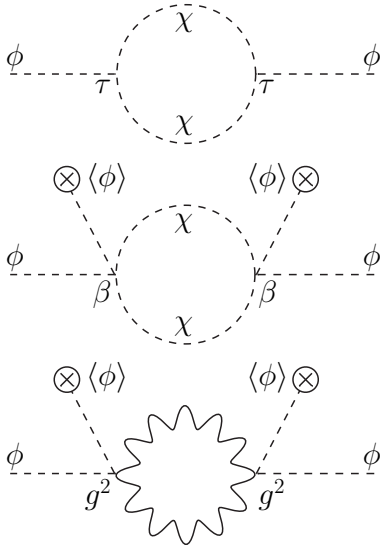}
\caption{\label{graphs} Typical one-loop diagrams that induce for
$\vev{\chi}=0$, $O(\tau/4\pi, \beta\vev{\phi}/4\pi, g^2\vev{\phi}/4\pi)$
renormalization to the mass of $45_H$ fields at the unification scale.
They are relevant for the PGB states, whose tree level mass is proportional to $a_2$.}
\end{figure}

The typical one-loop self energies, proportional to the $45_H$ VEVs,
are diagrammatically depicted in \fig{graphs}.
While the exchange of $16_{H}$ components is crucial,
the $\chi_R$ is not needed to obtain the large mass shifts.
In the phenomenologically allowed unification patterns it gives actually negligible contributions.

It is interesting to notice that the
$\tau$-induced mass corrections do not depend on
the gauge symmetry breaking, yielding an $SO(10)$
symmetric contribution to all scalars in $45_{H}$.

One is thus lead to the conclusion that any result based on the particular shape of the tree-level $45_H$ vacuum is drastically affected at the quantum level. Let us emphasize that although one may in principle avoid the $\tau$-term by means of e.g. an extra $Z_{2}$ symmetry, no symmetry can forbid the $\beta$-term and the gauge loop contributions.

In case one resorts to $126_H$, in place of $16_H$,
for the purpose of $B-L$ breaking,
the more complex tensor
structure of the class of $126_{H}^{\dagger}45_H^{2}126_{H}$ quartic invariants in the scalar potential may admit tree-level
contributions to the states $(1,3,0)$ and $(8,1,0)$ proportional to $\vev{126_H}$.
On the other hand, as mentioned above, whenever
$\vev{126_H}$ is small on the unification scale, the same considerations apply, as for the $16_H$ case.

%----------------------------------------------------------------
\subsection{The $\chi_{R}=0$ limit}
\label{sec:chi0limit}
%----------------------------------------------------------------

From the previous discussion it is clear that
the answer to the question whether the non-$SU(5)$ vacua are allowed at the quantum level is independent
on the specific value of the $B-L$ breaking VEV ($\chi_R\ll M_G$ in potentially realistic cases).

In order to simplify the study of the scalar potential beyond the classical level it is therefore convenient (and sufficient) to consider the $\chi_R = 0$ limit.

When $\chi_R =0$ the mass matrices of the $45_H$ and $16_H$ sectors
are not coupled.
The stationary equations in \eqs{eqstatmu}{eqstat0} lead to the
four solutions
\begin{itemize}
\item{$\omega = \omega_R = \omega_Y$\quad ($5\, 1_{Z}$)}
\item{$\omega = \omega_R = -\omega_Y$\quad ($5'\, 1_{Z'}$)}
\item{$\omega_R=0$ and $\omega_Y \neq 0$\quad ($3_c\, 2_L\, 2_R\, 1_{X}$)}
\item{$\omega_R \neq 0$ and $\omega_Y = 0$\quad ($4_C\, 2_L\, 1_R$)}
\end{itemize}
In what follows, we will focus our discussion on the last three cases only.

It is worth noting that the tree level spectrum in the $\chi_{R}=0$ limit is not
directly obtained from the general formulae given in \app{4516spectrum}, since \eq{eqstatnu} is trivially satisfied for $\chi_{R}=0$.
The corresponding scalar mass spectra are derived and discussed in \app{4516spectrumchi0}. Yet again, it is apparent that the non $SU(5)$ vacuum configurations exhibit unavoidable tachyonic states in the scalar spectrum.

%%%%%%%%%%%%%%%%%%%%%%%%%%%%%%%%%%%%%%%%%%%%%%%%%%%%%%%%%%%%%%%%
\section{The quantum vacuum}
\label{sec:quantumvacuum}
%%%%%%%%%%%%%%%%%%%%%%%%%%%%%%%%%%%%%%%%%%%%%%%%%%%%%%%%%%%%%%%%

%=========================================================================
\subsection{The one-loop effective potential}
\label{sec:1loopeffpot}
%=========================================================================

We shall compute the relevant one-loop corrections to the tree level
results by means of the one-loop effective potential (effective action
at zero momentum) \cite{Coleman:1973jx}.
We can formally write
\begin{equation}
V_{\rm eff}=V_{0}+\Delta V_s+\Delta V_f+\Delta V_g \, ,
\label{Veff}
\end{equation}
where $V_0$ is the tree level potential and $\Delta V_{s,f,g}$ denote the quantum contributions induced by scalars, fermions and gauge bosons respectively.
In dimensional regularization with the modified minimal subtraction
($\overline{MS}$) and in the Landau gauge, they are given by
\begin{multline}
\Delta V_s(\phi ,\chi ,\mu)= \\
\frac{\eta}{64\pi^2}
\Tr\left[W^4(\phi ,\chi)\left(\log\frac{W^2(\phi ,\chi)}{\mu^2}
-\frac{3}{2}\right)\right] \, ,
\end{multline}
\begin{multline}
\Delta V_f(\phi ,\chi ,\mu)= \\
-\frac{\kappa}{64\pi^2}
\Tr\left[M^4(\phi ,\chi)\left(\log\frac{M^2(\phi ,\chi)}{\mu^2}
-\frac{3}{2}\right)\right] \, ,
\end{multline}
\begin{multline}
\Delta V_g(\phi ,\chi ,\mu)= \\
\frac{3}{64\pi^2}
\Tr\left[\mathcal{M}^4(\phi ,\chi)\left(\log\frac{\mathcal{M}^2(\phi ,\chi)}{\mu^2}
-\frac{5}{6}\right)\right] \, ,
\end{multline}
with $\eta=1(2)$ for real (complex) scalars and $\kappa=2(4)$ for
Weyl (Dirac) fermions. $W$, $M$ and $\mathcal{M}$ are the functional scalar, fermion and gauge boson mass matrices respectively, as obtained from the tree level potential.

In the case at hand, we may write the functional scalar mass matrix,
$W^2(\phi,\chi)$ as a 77-dimensional hermitian matrix, with a lagrangian term
\beq
\label{defbasis}
\frac{1}{2}\psi^\dag W^2 \psi \, ,
\eeq
defined on the vector basis $\psi=(\phi,\chi,\chi^\ast)$. More explicitly, $W^{2}$ takes the block form
\beq
W^2(\phi,\chi)=\left(
\begin{array}{ccc}
V_{\phi\phi} & V_{\phi\chi} & V_{\phi\chi^\ast} \\
V_{\chi^{\ast}\phi} & V_{\chi^{\ast}\chi} & V_{\chi^{\ast}\chi^{\ast}} \\
V_{\chi\phi} & V_{\chi\chi} & V_{\chi\chi^{\ast}}
\end{array}
\right) \, ,
\label{W2matrix}
\eeq
where the subscripts denote the derivatives of the scalar potential with respect to the set of fields $\phi$, $\chi$ and $\chi^\ast$.
In the one-loop part of the effective potential $V\equiv V_0$.

We neglect the fermionic component of the effective potential since there are no fermions at the GUT scale (we assume that the right-handed (RH) neutrino mass is substantially lower than the unification scale).

The functional gauge boson
mass matrix, $\mathcal{M}^2(\phi ,\chi)$ is given in
\app{app:Treemasses},
\eqs{fielddepmass45}{fielddepmass16}.

%=========================================================================
\subsection{The one-loop stationary equations}
\label{sec:1loopstationary}
%=========================================================================

The first derivative of the one-loop part of the effective potential, with respect to the scalar field component $\psi_a$, reads
\begin{multline}
\label{der1deltaV}
\frac{\partial\Delta V_s}{\partial\psi_a}=
\frac{1}{64\pi^2}\Tr\left[\left\{W^2_{\psi_a},
  W^2\right\} \right. \\
\times \left(\log\frac{W^2}{\mu^2}-\frac{3}{2}\right)
+ \left. W^2W^2_{\psi_a}\right] \,
\end{multline}
where the symbol $W^{2}_{\psi_{a}}$ stands for the partial derivative of $W^{2}$ with respect to $\psi_{a}$.
Analogous formulae hold for $\partial\Delta V_{f,\, g}/\partial\psi_a$.
The trace properties ensure that \eq{der1deltaV} holds independently on whether $W^{2}$ does commute with its first derivatives or not.

The calculation of the loop corrected stationary equations
due to gauge bosons and scalar exchange is straightforward
(for $\chi_R=0$ the $45_H$ and $16_H$ blocks decouple in \eq{W2matrix}). On the other hand, the corrected equations are quite cumbersome and we do not explicitly report them here. It is enough to say that
the quantum analogue of \eq{eqstat0} admits analytically the same solutions as we had at the tree level. Namely, these are
$\omega_R = \omega_Y$, $\omega_R = -\omega_Y$, $\omega_R=0$ and $\omega_Y = 0$,
corresponding respectively to the
standard $5\, 1_Z$, flipped $5'\, 1_{Z'}$, $3_c\, 2_L\, 2_R\, 1_X$ and $4_C\, 2_L\, 1_R$ preserved subalgebras.

%=========================================================================
\subsection{The one-loop scalar mass}
\label{sec:1loopspectrum}
%=========================================================================

In order to calculate the second derivatives of the one-loop contributions to $V_{\rm eff}$ it is in general necessary to take into account the commutation properties of $W^{2}$ with its derivatives that enter as a series of nested commutators. The general expression can be written as
\begin{multline}
\frac{\partial^2\Delta V_s}{\partial\psi_a\partial\psi_b}=
\frac{1}{64\pi^2}\Tr \Big[ W^2_{\psi_a}W^2_{\psi_b}
+W^2W^2_{\psi_a\psi_b}  \\[1ex]
+\left[\left\{W^2_{\psi_a\psi_b}, W^2\right\}
+ \left\{W^2_{\psi_a},W^2_{\psi_b}\right\}\right]
\left(\log\frac{W^2}{\mu^2}-\frac{3}{2}\right) \\[1ex]
+  \sum_{m=1}^{\infty}(-1)^{m+1}\frac{1}{m}\sum_{k=1}^{m}{m\choose k}\left\{ W^{2},W^2_{\psi_a}\right\}  \\[1ex]
\times\left[W^{2},..\left[W^{2},W^2_{\psi_b}\right]..\right]\left(W^{2}-1\right)^{m-k}
\Bigr]
\label{der2deltaV}
\end{multline}
where the commutators in the last line are taken $k-1$ times.
Let us also remark that, although not apparent, the RHS of \eq{der2deltaV} can be shown to be symmetric under $a\leftrightarrow b$, as it should be. In specific cases (for instance when the nested commutators vanish or they can be rewritten as powers of a certain matrix commuting with $W$) the functional mass evaluated on the vacuum may take a closed form.

%-------------------------------------------------------------------------
\subsubsection{Running and pole mass}
\label{sec:polemass}
%-------------------------------------------------------------------------

The effective potential is a functional computed
at zero external momenta.
Whereas the stationary equations allow for the localization of the new
minimum (being the VEVs translationally invariant),
the mass shifts obtained from \eq{der2deltaV} define the
running masses
$\overline{m}_{ab}^2$
\beq
\overline{m}_{ab}^2 \equiv
\frac{\partial^2 V_{\rm eff}(\phi)}{\partial\psi_a\partial\psi_b}\Big |_{\vev{\psi}} =
m_{ab}^2 + \Sigma_{ab}(0)
\label{runningmass}
\eeq
where $m_{ab}^2$ are the renormalized masses and $\Sigma_{ab}(p^2)$
are the $\overline{MS}$ renormalized self-energies.
The physical (pole) masses $M_a^2$ are then obtained as a solution
to the equation
\beq
\mbox{det}\left[p^2\delta_{ab}-\left(\overline{m}_{ab}^2 +
%\Sigma_{ab}(p^2) - \Sigma_{ab}(0)
\Delta\Sigma_{ab}(p^2)
\right)\right] = 0
\label{invprop}
\eeq
where
\beq
\Delta\Sigma_{ab}(p^2) = \Sigma_{ab}(p^2) - \Sigma_{ab}(0)
\label{DeltaS}
\eeq
For a given eigenvalue
\beq
M_a^2 = \overline{m}_{a}^2
+ \Delta\Sigma_a (M_a^2)
\label{physmasses}
\eeq
gives the physical mass. The gauge and scheme dependence in \eq{runningmass} is canceled by the relevant contributions from \eq{DeltaS}. In particular, infrared
divergent terms in \eq{runningmass} related to the presence
of massless WGB in the Landau gauge cancel in \eq{physmasses}.

Of particular relevance is the case when $M_a$ is substantially smaller than
the (GUT-scale) mass of the particles that contribute to $\Sigma(0)$.
At $\mu=M_G$, in the $M_{a}^2 \ll M_G^2$ limit, one has
\beq
\Delta\Sigma_{a}(M_a^2) = O(M_a^4/M_G^2)\ .
\label{DeltaS-PGB}
\eeq
In this case the running mass computed from \eq{runningmass} contains the leading gauge independent corrections.
As a matter of fact, in order to study the vacua of the potential in \eq{Veff},
we need to compute the zero momentum mass corrections just to those states
that are tachyonic at the tree level and whose corrected mass turns out
to be of the order of $M_G/4\pi$.

We may safely neglect the one loop corrections
for all other states with masses of order $M_G$.
It is remarkable, as we shall see, that for $\chi_R=0$ the relevant corrections to the masses of the critical PGB states can be obtained from \eq{der2deltaV} with vanishing commutators.

%=========================================================================
\subsection{One-loop PGB masses}
\label{subsec:1Lmatching}
%=========================================================================

The stringent tree-level
constraint on the ratio $\omega_Y/\omega_R$, coming from the positivity
of the $(1,3,0)$ and $(8,1,0)$ masses, follows from the fact that some scalar masses depend only on the parameter $a_2$.
On the other hand, the discussion on the would-be global symmetries of the scalar potential shows that in general their mass should depend on other terms in the scalar potential, in particular $\tau$ and $\beta$.

A set of typical one-loop diagrams contributing $O(\vev{\phi}/4\pi)$ renormalization to the masses of $45_H$ states is depicted in \fig{graphs}. As we already pointed out the $16_H$ VEV does not play any role in the leading GUT scale corrections (just the interaction between $45_H$ and $16_H$, or with the massive gauge bosons is needed).
Therefore we henceforth work in the strict $\chi_R=0$ limit, that
simplifies substantially the calculation.
In this limit the scalar mass matrix in \eq{W2matrix} is block diagonal (c.f. \app{4516spectrumchi0}) and the leading corrections from the one-loop effective potential are encoded in the $V_{\chi^{\ast}\chi}$
sector.

More precisely, we are interested in the corrections to those
$45_H$ scalar states whose tree level mass depends only on $a_2$
and have the quantum numbers of the preserved non-abelian algebra
(see \sect{45a2} and \app{4516spectrumchi0}).
It turns out that focusing  to this set of PGB states
the functional mass matrix $W^2$ and its first derivative do commute for $\chi_R=0$  and \eq{der2deltaV} simplifies accordingly. This allows us to compute the relevant mass corrections in a closed form.

The calculation of the EP running mass
from \eq{der2deltaV} leads for the states $(1,3,0)$
and $(8,1,0)$ at $\mu=M_G$ to the mass shifts
\bwt
\bea
\label{310onthevac}
\Delta M^2(1,3,0)&=& \frac{1}{4\pi^2} \left[ \tau^2
+\beta^2(2\omega_R^2-\omega_R\omega_Y+2\omega_Y^2)
+g^4 \left(16 \omega _R^2+\omega _Y \omega _R+19 \omega _Y^2\right)\right] \, ,
\\[1ex]
\label{810onthevac}
\Delta M^2(8,1,0)&=& \frac{1}{4\pi^2} \left[ \tau^2
+\beta^2(\omega_R^2-\omega_R\omega_Y+3\omega_Y^2)
+g^4 \left(13 \omega _R^2+\omega _Y \omega _R+22 \omega _Y^2\right)\right]\, ,
\eea
\ewt
where the sub-leading (and gauge dependent) logarithmic terms
are not explicitly reported.
For the vacuum configurations of interest we find the results reported
in \app{app:1Lmasses}. In particular,  we obtain

\begin{itemize}
\item{
$\omega = \omega_R = -\omega_Y$\quad ($5'\, 1_{Z'}$):
\begin{align}
M^{2}(24,0) =
-4 a_2 \omega^2 +
\frac{\tau ^2 + (5 \beta ^2 + 34 g^4) \omega ^2}{4 \pi ^2}\,,
\label{m240}
\end{align}
}

\item{
$\omega_R=0$ and $\omega_Y \neq 0$\quad  ($3_c\, 2_L\, 2_R\, 1_X$):
\begin{multline}
M^{2}(1,3,1,0) = M^{2}(1,1,3,0) = \\
2 a_2 \omega_Y^2
+\frac{\tau ^2 + (2 \beta ^2 + 19 g^4) \omega _Y^2}{4 \pi ^2} \, ,
\label{m1310}
\end{multline}
\begin{align}
& M^{2}(8,1,1,0) = - 4 a_2 \omega_Y^2 +
\frac{\tau ^2 + (3 \beta ^2 + 22 g^4) \omega _Y^2}{4 \pi ^2}\,,
\label{m8110}
\end{align}
}

\item{
$\omega_R \neq 0$ and $\omega_Y = 0$\quad  ($4_C\, 2_L\, 1_R$):
\begin{align}
& M^{2}(1,3,0)  = - 4 a_2 \omega_R^2 +
\frac{\tau ^2 + (2 \beta ^2 + 16 g^4) \omega _R^2}{4 \pi ^2}  \label{m130}\,,
%\\[1ex]
\end{align}
\begin{align}
& M^{2}(15,1,0) = 2 a_2 \omega_R^2 +
\frac{\tau ^2 + (\beta ^2 + 13 g^4) \omega _R^2}{4 \pi ^2}\,.
\label{m1510}
\end{align}
}
\end{itemize}

In the effective theory language
\eqs{m240}{m1510}
can be interpreted as the one-loop GUT-scale matching due to the decoupling of the massive $SO(10)/G$ states where $G$ is the preserved gauge group.
These are the only relevant one-loop corrections needed in order
to discuss the vacuum structure of the model.

It is quite apparent that a consistent scalar mass spectrum can be obtained
in all cases, at variance with the tree level result.

In order to fully establish the existence of the non-$SU(5)$ minima at the quantum level one
should identify the regions of the parameter space supporting the desired vacuum configurations and estimate their depths. We shall address these issues in the next section.

%=========================================================================
\subsection{The one-loop vacuum structure}
\label{sec:1loopvacuum}
%=========================================================================

%-------------------------------------------------------------------------
\subsubsection{Existence of the new vacuum configurations}
\label{subsec:newvacua}
%-------------------------------------------------------------------------

The existence of the different minima of the one-loop effective potential is related to the values
of the parameters  $a_2$, $\beta$, $\tau$ and $g$ at the scale $\mu=M_G$. For the flipped $5'\, 1_{Z'}$ case it is sufficient,
as one expects, to assume the tree level condition $a_2<0$.
On the other hand, from \eqs{m1310}{m1510} we obtain
\begin{itemize}
\item{
$\omega_R=0$ and $\omega_Y \neq 0$\quad  ($3_c\, 2_L\, 2_R\, 1_X$):
\beq
%M^{2}(1,3,1,0) = M^{2}(1,1,3,0) = \\
- 8 \pi ^2 a_2 <
\frac{\tau ^2}{\omega _Y^2} + 2 \beta ^2 + 19 g^4 \, ,
\label{pm13loop}
\eeq
}
\item{
$\omega_R \neq 0$ and $\omega_Y = 0$\quad  ($4_C\, 2_L\, 1_R$):
\beq
%M^{2}(15,1,0) =
- 8 \pi ^2 a_2 <
\frac{\tau ^2}{\omega _R^2} + \beta ^2 + 13 g^4 \, .
\label{pm81loop}
\eeq
}
\end{itemize}

Considering for naturalness $\tau \sim \omega_{Y,R}$, \eqs{pm13loop}{pm81loop} imply $|a_2| < 10^{-2}$.
This constraint remains within the natural perturbative range for
dimensionless couplings.
While all PGB states whose mass is proportional to $-a_2$
receive large positive loop corrections,
quantum corrections are numerically irrelevant for all of the
states with GUT scale mass. On the same grounds we may safely neglect the multiplicative $a_2$ loop corrections induced by the $45_H$ states on the PGB masses.

%-------------------------------------------------------------------------
\subsubsection{Absolute minimum}
\label{subsec:absoluteminimum}
%-------------------------------------------------------------------------

It remains to show that the non $SU(5)$ solutions may actually be absolute minima of the potential.
To this end
it is necessary to consider the one-loop corrected stationary equations and calculate the vacuum energies in the relevant cases.
Studying the shape of the one-loop effective potential is a numerical
task. On the other hand, in the approximation of
neglecting at the GUT scale the logarithmic corrections, we
may reach non-detailed but definite conclusions.
For the three relevant vacuum configurations we obtain:
\bwt
\begin{itemize}
\item{
$\omega = \omega_R = -\omega_Y$\quad ($5'\, 1_{Z'}$)
\begin{align}
V(\omega, \chi_R=0) =
&-\frac{3 \nu ^4}{16 \pi ^2}
+ \left(\frac{5 \alpha  \nu ^2}{\pi ^2}+\frac{5 \beta  \nu ^2}{16
   \pi ^2}-\frac{5 \tau ^2}{16 \pi ^2}\right) \omega ^2  \nn \\[0.5ex]
&+ \left(-100 a_1-\frac{65 a_2}{4}+\frac{600 a_1^2}{\pi ^2}-\frac{45 a_1 a_2}{\pi ^2}-\frac{645 a_2^2}{32 \pi ^2}+\frac{100 \alpha
   ^2}{\pi ^2}+\frac{25 \alpha  \beta }{2 \pi ^2}+\frac{65 \beta ^2}{64 \pi ^2}-\frac{5 g^4}{2 \pi ^2}\right) \omega ^4 \, ,
\end{align}
}
\item{
$\omega_R=0$ and $\omega_Y \neq 0$\quad ($3_c\, 2_L\, 2_R\, 1_X$)
\begin{align}
V(\omega_Y, \chi_R=0) =
&-\frac{3 \nu ^4}{16 \pi ^2}
+ \left(\frac{3 \alpha  \nu ^2}{\pi ^2}+\frac{3 \beta  \nu
   ^2}{16 \pi ^2}-\frac{3 \tau ^2}{16 \pi ^2}\right) \omega _Y^2 \nn \\[0.5ex]
&+ \left(-36 a_1 -\frac{21 a_2}{4} +\frac{216 a_1^2}{\pi ^2}+\frac{33 a_1 a_2}{\pi ^2}+\frac{45 a_2^2}{32 \pi ^2}+\frac{36 \alpha
   ^2}{\pi ^2}+\frac{9 \alpha  \beta }{2 \pi ^2}+\frac{21 \beta ^2}{64 \pi ^2}-\frac{15 g^4}{16 \pi ^2}\right)\omega _Y^4 \, ,
\end{align}
}
\item{
$\omega_R \neq 0$ and $\omega_Y = 0$\quad ($4_C\, 2_L\, 1_R$)
\begin{align}
V(\omega_R, \chi_R=0) =
&-\frac{3 \nu ^4}{16 \pi ^2}
+\left(\frac{2 \alpha  \nu ^2}{\pi ^2}+\frac{\beta  \nu ^2}{8 \pi
   ^2}-\frac{\tau ^2}{8 \pi ^2}\right) \omega _R^2  \nn \\[0.5ex]
&+\left(-16 a_1-2 a_2+\frac{96 a_1^2}{\pi ^2}+\frac{42 a_1 a_2}{\pi ^2}+\frac{147 a_2^2}{32 \pi ^2}+\frac{16 \alpha ^2}{\pi
   ^2}+\frac{2 \alpha  \beta }{\pi ^2}+\frac{\beta ^2}{8 \pi ^2}-\frac{7 g^4}{16 \pi ^2}\right)\omega _R^4 \, .
\end{align}
}
\end{itemize}
\ewt

A simple numerical analysis reveals that for natural values of the dimensionless couplings and GUT mass parameters any of the qualitatively different vacuum configurations may be a global minimum of the one-loop effective potential in a large domain of the  parameter space.

This concludes the proof of existence of all of the group-theoretically
allowed vacua.
Nonsupersymmetric $SO(10)$ models
broken at $M_G$ by the $45_{H}$ SM preserving VEVs, do exhibit at the quantum level the full spectrum of intermediate symmetries. This is
crucially relevant for those chains that, allowed by gauge unification, are accidentally excluded by the tree level potential.

%-------------------------------------------------------------------------
\subsection{The extended survival hypothesis}
\label{subsec:ESH}
%-------------------------------------------------------------------------

In a realistic $SO(10)$ unification setup, throughout all the stages of the symmetry breaking one usually assumes that the scalar spectrum obeys the so
called extended survival hypothesis (ESH) that
reads~\cite{del Aguila:1980at}:
``at every stage of the symmetry breaking chain only those scalars
are present that develop a VEV
at the current or the subsequent levels of the spontaneous symmetry breaking''.

The ESH is equivalent to performing the minimal number of
fine-tunings imposed onto the scalar potential
%~\cite{Mohapatra:1982aq}
so that all the symmetry breaking steps are obtained at the desired scales.
On the technical side one must identify all the Higgs multiplets needed by the
breaking pattern and tune their mass
according to the gauge symmetry down to the scale of their VEVs.
The effects of the presence of these states at intermediate scales
has been considered in our recent analysis of non supersymmeteric $SO(10)$ unification patterns~\cite{Bertolini:2009qj},
up to one exception that we shall now shortly comment upon.

The relevant patterns preserve in the first stage
the group $3_c\, 2_L\, 2_R\, 1_X$
(for $\omega_R = 0$) and $4_C\, 2_L\, 1_R$ (for $\omega_Y = 0$).
The breaking to the SM gauge group $3_c\, 2_L\, 1_Y$ is achieved by means of the VEV $\chi_R$, constrained to stay at an intermediate scale by gauge unification. Minimally, one must therefore maintain at this scale either of the $16_H$ multiplets $(1,1,2,\tfrac{1}{2})$
and $(\overline{4},1,-\tfrac{1}{2})$,
in the $3_c\, 2_L\, 2_R\, 1_X$
and $4_C\, 2_L\, 1_R$ cases respectively.

As one can see from the scalar spectrum given in \app{4516spectrumchi0}, in the $3_c\, 2_L\, 2_R\, 1_X$
vacuum, the scalars $(1,2,1,\tfrac{1}{2})$ and $(1,1,2,-\tfrac{1}{2})$ receive a mass contribution that is linear
in the D-odd VEV $\omega_Y$ and that breaks their degeneracy.
Thus, just the RH doublet $(1,1,2,\tfrac{1}{2})$, which contains the field acquiring the VEV $\chi_R$, may be minimally fine-tuned at that mass scale.

Turning on $\omega_Y \neq 0$ or $\omega_R \neq 0$ at the $\chi_R$
scale leads to a non-minimal set of Higgs states at the intermediate scale~\cite{Bertolini:2009qj}, namely the $45_H$ multiplets $(1,1,3,0)$ and (15,1,0) in the $3_c\, 2_L\, 2_R\, 1_X$ and
in the $4_C\, 2_L\, 1_R$ setting respectively (these are the accidentally tachyonic states at the tree level).
Inspection of the one-loop mass spectra shows that the needed minimal fine-tuning can be indeed performed.

It is worth noticing that although in the $3_c\, 2_L\, 2_R\, 1_X$ stage D-parity is broken by $\omega_Y$, the masses
of the states $(1,3,1,0)$ and $(1,1,3,0)$, depending quadratically on $\omega_Y$ (see \app{4516spectrumchi0}), do
remain degenerate and are {\em both} tuned at the
scale $\omega_R\sim \chi_R$, where the LR symmetry is broken.
The presence of the additional LH triplet $(1,3,1,0)$ at the intermediate scale $\omega_R$ has a welcome impact on the gauge coupling running. Compared to the results given in \cite{Bertolini:2009qj} for such a breaking pattern, the intermediate $B-L$ scale is raised by
almost one order of magnitude (to about $10^{11}$ GeV), while the GUT scale is slightly lowered to about $10^{16}$ GeV.
Detailed thresholds effects can be considered once the model dependent scalar spectrum is fully worked out.

A final comment is in order.
All of the states exchanged in the relevant mass loop corrections in \sect{subsec:1Lmatching} have natural GUT masses. On the other hand, the ESH requires tuning the masses of some of these states
at a much lower scale. In a realistic setting, this involves some of the $16_H$ submultiplets. The fine tuning
apparently generates an infrared divergence problem in the
one-loop corrections. However,
in analogy to our discussion of the WGB contributions to the effective potential in \sect{sec:1loopspectrum}, the infrared divergent terms appearing
in the one-loop zero momentum mass corrections disappear when
considering the corrections to the physical pole masses. Thus, they can be safely discarded.

%%%%%%%%%%%%%%%%%%%%%%%%%%%%%%%%%%%%%%%%%%%%%%%%%%%%%%%%%%%%%%%%%
\section{Summary and outlook}
\label{sec:outlook}
%%%%%%%%%%%%%%%%%%%%%%%%%%%%%%%%%%%%%%%%%%%%%%%%%%%%%%%%%%%%%%%%%

In this paper, we have scrutinized the longstanding result that the class of the minimal nonsupersymmetric $SO(10)$ unified models, with the GUT symmetry broken by the VEVs of the 45-dimensional adjoint representation, cannot provide a successful gauge unification. This common knowledge was based on the observation that the tree level dynamics of the minimal scalar sector allowed only for ``$SU(5)$'' breaking patterns. This, in turn, clashes with the
intermediate symmetries required by nonsupersymmetric $SO(10)$ unification, that enforce an intermediate threshold well below the GUT scale.

We argued that the old result is an artifact of the tree level
Higgs potential and showed that quantum corrections have a dramatic impact. The minimization of the one-loop effective potential
in the paradigmatic $\chi_R=0$ limit shows
that the simplest $SO(10)$ model with a $45_H\oplus 16_{H}$ Higgs sector allows for any of the intermediate symmetry patterns available to the pair of the SM-preserving VEVs in $45_{H}$.
In particular, the $SU(4)_C \otimes SU(2)_L \otimes U(1)_R$
and
$SU(3)_c \otimes SU(2)_L \otimes SU(2)_R \otimes U(1)_{B-L}$ chains are supported.
Our result generally applies to any Higgs sector where the vacuum is dominated
by the $45_H$ VEVs.

This observation opens the option of reconsidering the minimal nonsupersymmetric $SO(10)$ model as a reference framework for model building.
Extending the Higgs sector to include one $10_{H}$ (together with either one ${126}_H$ or one ${16}_H$) provides the playground
for exploring the possibility of a realistic and predictive GUT,
along the lines of the recent efforts in the supersymmetric context.

%%%%%%%%%%%%%%%%%%%%%%%%%%%%%%%%%%%%%%%%%%%%%%%%%%%%%%%%%%%%%%%%%

\subsection*{Acknowledgments}

S.B. acknowledges support by the MRTN European Program
MRTN-CT-2006-035863. The work of M.M. is supported by the Royal Institute of Technology (KTH), Contract No. SII-56510.

%\subsection*{Note added}

\clearpage
%%%%%%%%%%%%%%%%%%%%%%%%%
\appendix
%%%%%%%%%%%%%%%%%%%%%%%%%

%%%%%%%%%%%%%%%%%%%%%%%%%%%%%%%%%%%%%%%%%%%%%%%%%%%%%%%%%%%%%%%%%
\section{$SO(10)$ algebra representations}
\label{app:so10algebra}
%%%%%%%%%%%%%%%%%%%%%%%%%%%%%%%%%%%%%%%%%%%%%%%%%%%%%%%%%%%%%%%%%

We briefly recall here for convenience the basics of $SO(10)$ algebra representations.
For a general introduction see Refs.~\cite{Slansky:1981yr,Georgi:1982jb}.

\subsection{Tensorial representations}
\label{app:tensorrep}

The hermitian and antisymmetric generators of the fundamental representation of $SO(10)$ are given by
\begin{equation}
(\epsilon_{ij})_{ab}=-i(\delta_{a[i}\delta_{bj]}) \, ,
\end{equation}
where $a,b,i,j=1,..,10$ and the square bracket stands for anti-symmetrization.
They satisfy the $SO(10)$ commutation relations
\beq
\label{so10commrel}
\left[ \epsilon_{ij}, \epsilon_{kl} \right] = -i ( \delta_{jk}\epsilon_{il} - \delta_{ik}\epsilon_{jl} - \delta_{jl}\epsilon_{ik} + \delta_{il}\epsilon_{jk}) \, ,
\eeq
with normalization
\beq
\Tr \epsilon_{ij}\epsilon_{kl} = 2\ \delta_{i[k}\delta_{jl]} \,
\eeq
and Dynkin index 2.

The fundamental (vector) representation $\phi_a$ $ (a=1,...,10)$ transforms
as
\beq
\label{10transf}
\phi_a \rightarrow \phi_a - \frac{i}{2} \lambda_{ij} (\epsilon_{ij}\phi)_a \, ,
\eeq
where $\lambda_{ij}$ are the infinitesimal parameters of the transformation.

The adjoint representation is then obtained as the
antisymmetric part of the 2-index $10_a\otimes 10_b$ tensor $\phi_{ab}$ $(a,b=1,..,10)$
and transforms as
\beq
\phi_{ab}\rightarrow  \phi_{ab}-\frac{i}{2}\lambda_{ij}\left[\epsilon_{ij},\phi\right]_{ab} \, .
\eeq
Notice that
$\left[\epsilon_{ij},\phi\right]^T = - \left[\epsilon_{ij},\phi\right]$
and $\left[\epsilon_{ij},\phi\right]^\dag = \left[\epsilon_{ij},\phi\right]$.

\subsection{Spinorial representations}
\label{app:spinorrep}

Following the notation of Ref.~\cite{Babu:1984mz}, the $SO(10)$ generators $S_{ij}$ ($i,j=0,..,9$) acting on the 32-dimensional
spinor $\Xi$ are defined as
\begin{equation}
\label{spingen32}
S_{ij}=\frac{1}{4i}\left[\Gamma_i,\Gamma_j\right] \, , \ \
\left\{\Gamma_i,\Gamma_j\right\}=2\delta_{ij} \, ,
\end{equation}
with an explicit representation given by % \cite{Georgi:1979dq}
\begin{equation}
\Gamma_0=\left(
\begin{array}{cc}
 0 & I_{16}  \\
 I_{16} & 0
\end{array}
\right) \, , \ \
\Gamma_p=\left(
\begin{array}{cc}
 0     & is_p  \\
 -is_p & 0
\end{array}
\right) \, , \  p=1,...,9 \; ,
\label{cliffalgebra}
\end{equation}
where the $s_p$ matrices are defined as ($k=1,..,3$)
\begin{equation}
s_k=\eta_k\rho_3 \, , \quad s_{k+3}=\sigma_k\rho_1 \, , \quad s_{k+6}=\tau_k\rho_2 \, .
\end{equation}
The matrices $\sigma_k$, $\tau_k$, $\eta_k$ and $\rho_k$,
are given by the following tensor products of $2\times 2$ matrices
\begin{align}
\label{sigmatauetaro}
\sigma_k&=I_{2}\otimes I_{2}\otimes I_{2}\otimes \Sigma_k \, , \nn\\
\tau_k  &=I_{2}\otimes I_{2}\otimes \Sigma_k\otimes I_{2} \, , \\
\eta_k  &=I_{2}\otimes \Sigma_k\otimes I_{2}\otimes I_{2} \, , \nn\\
\rho_k  &=\Sigma_k\otimes I_{2}\otimes I_{2}\otimes I_{2} \, \nn,
\end{align}
where $\Sigma_k$ stand for the ordinary Pauli matrices.
Defining
\begin{equation}
s_{pq}=\frac{1}{2i}\left[s_p,s_q\right]
\end{equation}
for $p,q=1,..,9$, the algebra (\ref{spingen32}) is represented by
\begin{equation}
S_{p0}=\frac{1}{2}\left(
\begin{array}{cc}
 s_p & 0  \\
 0 & -s_p
\end{array}
\right)
\, , \ \
S_{pq}=\frac{1}{2}\left(
\begin{array}{cc}
 s_{pq} & 0  \\
 0 & s_{pq}
\end{array}
\right) \, .
\label{32spingen}
\end{equation}
The Cartan subalgebra is spanned over $S_{03}$, $S_{12}$, $S_{45}$, $S_{78}$ and $S_{69}$.
One can construct a chiral projector $\Gamma_{\chi}$, that splits the 32-dimensional spinor $\Xi$ into a pair of irreducible 16-dimensional components:
 \begin{equation}
\Gamma_{\chi}=2^{-5} S_{03}S_{12}S_{45}S_{78}S_{69}=\left(
\begin{array}{cc}
 -I_{16} & 0  \\
 0 & I_{16}
\end{array}
\right)
\, .
\end{equation}
It is readily verified that $\Gamma_\chi$ has the following properties:
$\Gamma_\chi^2=I_{32}$, $\left\{ \Gamma_\chi, \Gamma_i \right\}=0$ and hence
$\left[ \Gamma_\chi, S_{ij} \right]=0$.
Introducing the chiral projectors $P_\pm=\frac{1}{2}(I_{32}\mp\Gamma_{\chi})$,
the irreducible chiral spinors are defined as
\beq
\chi_+ = P_+\Xi\equiv
\left(
\begin{array}{l}
 \chi  \\
 0
\end{array}
\right) \, , \quad
\chi_- = P_-\Xi\equiv
\left(
\begin{array}{l}
 0  \\
 \chi^{c}
\end{array}
\right) \, ,
\eeq
where $\chi^{c}\equiv C\chi^{\ast}$ and $C$ is the $SO(10)$ charge conjugation matrix (see next subsection).
Analogously, we can use the chiral projectors to write $S_{ij}$ as
\beq
\label{genprojected}
S_{ij}=P_+S_{ij}P_+ + P_-S_{ij}P_-
\equiv
\frac{1}{2}
\left(
\begin{array}{cc}
 \sigma_{ij} & 0  \\
 0 & \tilde{\sigma}_{ij}
\end{array}
\right) \, ,
\eeq
where the properties $\left[ P_\pm, S_{ij} \right]=0$,
$P_\pm^2=P_\pm$ and $P_+ + P_- = I_{32}$ were used.

Finally, matching \eq{genprojected} with \eq{32spingen}, one
identifies the hermitian generators $\sigma_{ij}/2$ and $\tilde{\sigma}_{ij}/2$ acting
on the $\chi$ and $\chi^c$ spinors, respectively, as
\beq
\sigma_{p0} = s_p\;,\ \sigma_{pq}=s_{pq} \;,\
\tilde{\sigma}_{p0} = -s_p\;,\ \tilde{\sigma}_{pq}=s_{pq} \, .
\eeq
From their normalization
\beq
\label{Dynkinspinor}
\tfrac{1}{4}\Tr \sigma_{ij}\sigma_{kl} =
\tfrac{1}{4}\Tr \tilde{\sigma}_{ij}\tilde{\sigma}_{kl} =
4\ \delta_{i[k}\delta_{jl]} \, ,
\eeq
we recover the Dynkin index 4 of the 16-dimensional spinorial representation.

It is convenient to trace out the $\sigma$-matrices in the invariants built off the adjoint representation in the natural basis $\Phi\equiv\sigma_{ij}\phi_{ij}/4$.
From the traces of two and four $\sigma$-matrices one obtains
\begin{align}
\label{tr2sig}
&\Tr\Phi^2= - 2 \Tr\phi^2 \, , \\[1ex]
%\end{equation}
%and
%\begin{equation}
\label{tr4sig}
&\Tr\Phi^4= \tfrac{3}{4} \left(\Tr\phi^2\right)^2 - \Tr\phi^4
\, .
\end{align}

In order to maintain a consistent notation, from now on
we shall label the indices of the spinorial generators from
1 to 10, and use the following mapping from the basis of Ref.~\cite{Babu:1984mz} into the basis of Ref.~\cite{Bajc:2004xe}
for both vectors and tensors:
$\{0 3 1 2 4 5 7 8 6 9\} \rightarrow \{1 2 3 4 5 6 7 8 9 10\}$.

\subsection{The charge conjugation $C$}
\label{app:spinorandC}

%We construct an explicit representation for the charge conjugation operator $C$.

According to the notation of the previous subsection,
the spinor $\chi$ and its complex conjugate $\chi^{\ast}$ transform as
\beq
\chi\rightarrow \chi - \frac{i}{4} \lambda_{ij}\sigma_{ij} \chi \, , \quad
\chi^{\ast}\rightarrow \chi^{\ast} +\frac{i}{4} \lambda_{ij}\sigma_{ij}^{T} \chi^{\ast} \, .
\eeq
The charge conjugated spinor $\chi^{c}\equiv C\chi^{\ast}$ obeys
\beq
\chi^{c}\rightarrow \chi^{c} - \frac{i}{4} \lambda_{ij}\tilde{\sigma}_{ij} \chi^{c} \, ,
\eeq
 and thus $C$ satisfies
\beq
\label{Cproperty}
C^{-1}\tilde{\sigma}_{ij}C=-\sigma_{ij}^{T} \, .
\eeq
Taking into account \eq{sigmatauetaro}, a formal solution reads
\beq
\label{Crelations}
C=\sigma_2\tau_2\eta_2\rho_2 \, ,
\eeq
which in our basis yields
\begin{multline}
C = \text{antidiag}(+1,-1,-1,+1,-1,+1,+1,-1, \\
                    -1,+1,+1,-1,+1,-1,-1,+1) ,
\end{multline}
 and hence $
C=C^{\ast}=C^{-1}=C^{T}=C^{\dag}$.

\subsection{The Cartan generators}
\label{app:explicitgenerators}

It is convenient to write the five $SO(10)$ Cartan generators in the $3_c\, 2_L\, 2_R\, 1_X$ { basis} ($X=(B-L)/2$),
where { the physical interpretation is obvious. For} the spinorial
representation we have
\bea
& T^{(3)}_R=\tfrac{1}{4}(\sigma_{12}+\sigma_{34}) \, , \quad
\widetilde{T}^{(3)}_R=\tfrac{1}{4}(-\sigma_{12}+\sigma_{34}) \, , &
\nn\\
&T^{(3)}_L=\tfrac{1}{4}(\sigma_{34}-\sigma_{12}) \, , \quad
\widetilde{T}^{(3)}_L=\tfrac{1}{4}(\sigma_{34}+\sigma_{12}) \, , &
\nn\\
& T^{(3)}_c=\widetilde{T}^{(3)}_c=\tfrac{1}{4}(\sigma_{56}-\sigma_{78}) \nn \, ,
\\
&T^{(8)}_c=\widetilde{T}^{(8)}_c=\tfrac{1}{4\sqrt{3}}(\sigma_{56}+\sigma_{78}-2\sigma_{910}) \, ,&
\nn\\
&T_X=\widetilde{T}_X=-\tfrac{2}{3}(\sigma_{56}+\sigma_{78}+\sigma_{910}) \, .&
\label{cartanT}
\eea

While the $T$'s act on $\chi$, the $\widetilde{T}$'s (characterized by a sign flip in $\sigma_{1i}$)
act on $\chi^c$.
The normalization of the Cartan generators is chosen according to the usual SM convention.
A { GUT-consistent normalization across} all generators is obtained by rescaling $T_X$ (and $\widetilde{T}_X$) by $\sqrt{3/2}$.

In order to obtain the physical generators acting on the fundamental representation it is enough to replace
{ $\sigma_{ij}/2$ in \eq{cartanT} by $\epsilon_{ij}$.}

{ With this information at hand}, one can identify the spinor components of { $\chi$ and $\chi^{c}$}
\begin{multline}
\label{chiembedding}
\chi = (\nu,u_1,u_2,u_3,l,d_1,d_2,d_3, \\
           -d^c_3,d^c_2,d^c_1,-l^c,u^c_3,-u^c_2,-u^c_1,\nu^c) \, ,
\end{multline}
and
\begin{multline}
\label{chicembedding}
{\chi^c} = (\nu^c,u^c_1,u^c_2,u^c_3,l^c,d^c_1,d^c_2,d^c_3, \\
               -d_3,d_2,d_1,-l,u_3,-u_2,-u_1,\nu)^{\ast} \, ,
\end{multline}
{ where a self-explanatory SM notation has been naturally extended into the scalar sector}.
In particular, the relative signs in \eqs{chiembedding}{chicembedding} arise from the charge conjugation of the
$SO(6)\sim SU(4)_C$ and $SO(4)\sim SU(2)_L\otimes SU(2)_R$
components of $\chi$ and $\chi^c$.

The standard and flipped embeddings of $SU(5)$
%mentioned in \sect{sec:su5vsflippedsu5}
commute with two different Cartan generators, $Z$ and $Z'$ respectively:
\beq
Z =-4T^{(3)}_R + 6T_X \, , \quad
Z' =4T^{(3)}_R + 6T_X \, .
\eeq
Given the relation $\Tr (T^{(3)}_R)^2= \frac{3}{2}\Tr T_X^2$
one obtains
\beq
\Tr (YZ)=0 \, , \;
\Tr (YZ')\neq 0 \, ,
\eeq
where $Y = T^{(3)}_R + T_X$
is the weak hypercharge generator.

As a consequence, the standard $SU(5)$ contains the SM group,
while $SU(5)'$ has a subgroup
$SU(3)_c\otimes SU(2)_L\otimes U(1)_{Y'}$, with
\beq
\label{defZ}
Y' =-T^{(3)}_R+T_X \, .
\eeq
In terms of $Z'$ and of $Y'$ the weak hypercharge reads
\beq
\label{deffwhyp}
Y=\tfrac{1}{5}(Z'-Y') \, .
\eeq
Using the explicit form of the Cartan generators in the vector
representation one finds
\begin{align}
Z'& \propto \text{diag}(-1,-1,+1,+1,+1)\otimes \Sigma_2 \, ,\\[1ex]
Z &\propto \text{diag}(+1,+1,+1,+1,+1)\otimes \Sigma_2 \, .
\end{align}
{ The vacuum configurations
$\omega_R = -\omega_Y$ and $\omega_R = \omega_Y$ in \eq{vacua}
are aligned} with the $Z'$ and the $Z$ generator respectively,
thus preserving $SU(5)'\otimes U(1)_{Z'}$ and $SU(5)\otimes U(1)_{Z}$, respectively.

\clearpage
%%%%%%%%%%%%%%%%%%%%%%%%%%%%%%%%%%%%%%%%%%%%%%%%%%%%%%%%%%%%%%
\section{Vacuum stability}
%%%%%%%%%%%%%%%%%%%%%%%%%%%%%%%%%%%%%%%%%%%%%%%%%%%%%%%%%%%%%%
\label{boundedpot}

The boundedness of the scalar potential is needed
in order to ensure the global stability of the vacuum.
The requirement that the { potential is bounded} from below sets non trivial constraints
on the quartic interactions.
We do not { provide a fully general analysis for the whole field space,
but limit ourselves to the constraints obtained for
the given vacuum directions}.

%%%%%%%%%%%%%%%%%%%%%%%%%%%%%%%%%%%%%%%%%%%%%%%%%%%%%%%%%%%%%%
\subsection{$(\omega_R$, $\omega_Y$, $\chi_R) \neq 0$}
%%%%%%%%%%%%%%%%%%%%%%%%%%%%%%%%%%%%%%%%%%%%%%%%%%%%%%%%%%%%%%

From the quartic part of the scalar potential $V_0^{(4)}$
one obtains

\begin{align}
& 4a_1(2\omega_R^2 +3\omega_Y^2)^2
 +\frac{a_2}{4}(8\omega_R^4 + 21\omega_Y^4 + 36\omega_R^2\omega_Y^2) \nn \\[1ex]
& + \frac{\lambda_1}{4}\chi_R^4
+ 4\alpha\chi_R^2(2\omega_R^2 +3\omega_Y^2)\nn \\[1ex]
& +\frac{\beta}{4}\chi_R^2(2\omega_R + 3\omega_Y)^2
 -\frac{\tau}{2}\chi_R^2(2\omega_R + 3\omega_Y) > 0
\end{align}
Notice that the $\lambda_2$ term vanishes along the
$16_H$ vacuum direction.

%%%%%%%%%%%%%%%%%%%%%%%%%%%%%%%%%%%%%%%%%%%%%%%%%%%%%%%%%%%%%%
\subsection{$\omega_R = \omega_Y=0$, $\chi_R\neq 0$}
%%%%%%%%%%%%%%%%%%%%%%%%%%%%%%%%%%%%%%%%%%%%%%%%%%%%%%%%%%%%%%
Along this direction the quartic potential  { $V_0^{(4)}$ reads}

\beq
\label{V4only16}
V_0^{(4)}=\tfrac{1}{4} \lambda _1 \chi _R^4 \, ,
\eeq
which implies

\beq
\label{boundonly16}
\lambda_1 > 0 \, .
\eeq

%%%%%%%%%%%%%%%%%%%%%%%%%%%%%%%%%%%%%%%%%%%%%%%%%%%%%%%%%%%%%%
\subsection{$\omega = \omega_R = -\omega_Y$, $\chi_R=0$}
%%%%%%%%%%%%%%%%%%%%%%%%%%%%%%%%%%%%%%%%%%%%%%%%%%%%%%%%%%%%%%

{ From now on, we focus on the $\chi_R=0$ case, c.f. \sect{sec:chi0limit}}.
On this { orbit} the quartic part of the scalar potential reads

\beq
\label{V4only45}
V_0^{(4)}=\tfrac{5}{4} \omega ^4 (80 a_1+13 a_2) \, .
\eeq

Taking into account that the scalar mass spectrum implies
$a_2<0$, we obtain

\beq
\label{bound45only}
a_1 > -\tfrac{13}{80} a_2 \, .
\eeq

%%%%%%%%%%%%%%%%%%%%%%%%%%%%%%%%%%%%%%%%%%%%%%%%%%%%%%%%%%%%%%
\subsection{$\omega_R=0$, $\omega_Y \neq 0$, $\chi_R=0$}
%%%%%%%%%%%%%%%%%%%%%%%%%%%%%%%%%%%%%%%%%%%%%%%%%%%%%%%%%%%%%%

At the tree level this VEV configuration does not correspond
to a minimum of the potential.
It is nevertheless useful to inspect the stability conditions along this direction.
Since

\beq
\label{V4-3211}
V_0^{(4)}=  \tfrac{3}{4} (48 a_1+7 a_2) \omega _Y^4 \ ,
\eeq
boundedness is obtained, independently on the sign of $a_2$, when

\beq
\label{boundhierarch}
a_1 > - \tfrac{7}{48} a_2 \ .
\eeq

%%%%%%%%%%%%%%%%%%%%%%%%%%%%%%%%%%%%%%%%%%%%%%%%%%%%%%%%%%%%%%
\subsection{$\omega_R \neq 0$, $\omega_Y = 0$, $\chi_R=0$}
%%%%%%%%%%%%%%%%%%%%%%%%%%%%%%%%%%%%%%%%%%%%%%%%%%%%%%%%%%%%%%

In analogy with the previous case we have

\beq
V_0^{(4)}= 2 (8 a_1+a_2) \omega _R^4
\, ,
\label{V4-421}
\eeq
which implies the constraint

\beq
a_1 > - \tfrac{1}{8} a_2
\, .
\eeq

In the case $a_2<0$ the constraint in \eq{bound45only} { provides the global lower bound on $a_1$.}

%\clearpage

%%%%%%%%%%%%%%%%%%%%%%%%%%%%%%%%%%%%%%%%%%%%%%%%%%%%%%%%%%%%%%%%%
\section{Tree level mass spectra}
%%%%%%%%%%%%%%%%%%%%%%%%%%%%%%%%%%%%%%%%%%%%%%%%%%%%%%%%%%%%%%%%%
\label{app:Treemasses}

\subsection{Gauge bosons}
\label{app:gaugespectrum}
From the scalar kinetic terms
\beq
\label{gauge45piece}
\tfrac{1}{4}\Tr(D_{\mu}\phi)^{\dag}(D^{\mu}\phi) \, ,
\eeq
and
\beq
\label{gauge16piece}
\tfrac{1}{2}(D_{\mu}\chi)^{\dag} (D^{\mu}\chi) \,
+ \tfrac{1}{2}(D_{\mu}\chi^c)^{\dag} (D^{\mu}\chi^c) \, ,
\eeq
one may write the field dependent mass matrices for the gauge bosons
as
\begin{align}
\label{fielddepmass45}
\mathcal{M}_A^2(\phi)_{(ij)(kl)}
&= \frac{g^2}{2} \Tr [\epsilon_{(ij)}, \phi][\epsilon_{(kl)}, \phi]     \, ,
%&=-g^2(2\phi_{[jk}\phi_{li]}+\delta_{j[l}\phi^2_{k]i}+\delta_{i[k}\phi^2_{l]j}) \, ,
\\[0ex]
\label{fielddepmass16}
\mathcal{M}_A^2(\chi)_{(ij)(kl)}
&=\frac{g^2}{4} \chi^\dag\{\sigma_{(ij)},\sigma_{(kl)}\}\chi \, .
\end{align}
where $(ij),\, (kl)$ { stand for ordered pairs of indices, and $\epsilon_{ij}$ ($\sigma_{ij}/2$) with $i,j=1,..,10$
are the generators of the fundamental (spinor)} representation (see \app{app:so10algebra}).

{
\eqs{fielddepmass45}{fielddepmass16}, evaluated on the generic ($\omega_{R,Y}\neq 0,\ \chi_{R}\neq 0$) vacuum, yield
the following contributions to the tree level gauge boson masses}:

\subsubsection{Gauge bosons masses from 45}
\label{gaugespectrum45}
{ Focusing on \eq{gauge45piece} one obtains}

\begin{align}
&\mathcal{M}_A^2(1,1,+1)=4 g^2 \omega _R^2 \, ,
\nn \\[0ex]
&\mathcal{M}_A^2(\overline{3},1,-\tfrac{2}{3})=4 g^2 \omega _Y^2 \, ,
\nn\\[0ex]
&\mathcal{M}_A^2(1,3,0)=0 \, ,
\nn\\[0ex]
&\mathcal{M}_A^2(8,1,0)=0 \, ,
\\[0ex]
&\mathcal{M}_A^2(3,2,-\tfrac{5}{6})=g^2 \left(\omega _R-\omega _Y\right)^2 \, ,
\nn\\[0ex]
&\mathcal{M}_A^2(3,2,+\tfrac{1}{6})=g^2 \left(\omega _R+\omega _Y\right)^2 \, ,
\nn\\[0ex]
&\mathcal{M}_A^2(1,1,0)=
\left(
\begin{array}{cc}
0 & 0 \\
0 & 0
\end{array}
\right) \, ,\nn
\end{align}
where the SM singlet matrix is defined on the basis ($\psi^{45}_{24}$, $\psi^{45}_{1}$),
with the superscript referring to the original $SO(10)$ representation and
the subscript to the standard $SU(5)$ embedding (see \Table{tab:45decomp}).

Note that, in the limits of standard $5\, 1_{Z}$ $(\omega_R = \omega_Y)$, flipped $5'\, 1_{Z'}$ $(\omega_R = -\omega_Y)$,
$3_c\, 2_L\, 2_R\, 1_X$ ($\omega_R=0$) and $4_C\, 2_L\, 1_R$ ($\omega_Y=0$) vacua, we have respectively
25, 25, 15 and 19 massless gauge bosons, { as expected}.

\subsubsection{Gauge bosons masses from 16}
{ The contributions from \eq{gauge16piece} read}

\begin{align}
&\mathcal{M}_A^2(1,1,+1)=g^2 \chi _R^2 \, ,
 \nn\\[0ex]
&\mathcal{M}_A^2(\overline{3},1,-\tfrac{2}{3})=g^2 \chi _R^2 \, ,
\nn\\[0ex]
&\mathcal{M}_A^2(1,3,0)=0 \, ,
\nn\\[0ex]
&\mathcal{M}_A^2(8,1,0)=0 \, ,
\\[0ex]
&\mathcal{M}_A^2(3,2,-\tfrac{5}{6})=0 \, ,
\nn\\[0ex]
&\mathcal{M}_A^2(3,2,+\tfrac{1}{6})=g^2 \chi _R^2 \, ,
\nn\\[0ex]
&\mathcal{M}_A^2(1,1,0)=
\left(
\begin{array}{cc}
 \frac{3}{2}  & \sqrt{\frac{3}{2}}  \\
 \sqrt{\frac{3}{2}}  & 1
\end{array}
\right)g^2 \chi _R^2 \, ,
\nn
\end{align}
where the last matrix is { again} spanned over ($\psi^{45}_{24}$, $\psi^{45}_{1}$), yielding

\begin{align}
\text{Det} \mathcal{M}_A^2(1,1,0)&=0 \, , \\[0ex]
\Tr \mathcal{M}_A^2(1,1,0)&=\tfrac{5}{2} g^2 \chi _R^2 \, .
\end{align}

The number of vanishing entries corresponds to the dimension
of the $SU(5)$ algebra
preserved by the $16_H$ VEV $\chi_R$.

Summing together the $45_H$ and $16_H$ contributions, we recognize 12 massless states, { that correspond} to the SM gauge bosons.

\subsection{Anatomy of the scalar spectrum}
\label{app:scalarspectrum}

In order to understand the dependence of the scalar masses on
the various parameters in the Higgs potential
we detail the scalar mass spectrum in the relevant limits of the scalar couplings, according to
the discussion on the accidental global symmetries in \sect{sect:understanding}.

\subsubsection{45 only}
\label{45only}

Applying the stationary conditions in \eqs{eqstatmu}{eqstat0},
to the flipped $5'\, 1_{Z'}$ vacuum with $\omega=\omega_R=-\omega_Y$, we find
\begin{align}
&M^2(24,0)= -4 a_2 \omega ^2 \, , \nn\\[0.5ex]
&M^2(10,-4) = 0 \, , \\[0.5ex]
&M^2(1,0) = 2 \left(80 a_1 + 13 a_2\right) \omega ^2 \, ,\nn
\end{align}
and, as expected, the spectrum exhibits 20 WGB and 24 PGB whose
mass depends on $a_2$ only.
The required positivity of the scalar masses gives the constraints
\beq
\label{fsu5only45cond}
a_2 < 0 \qquad \text{and} \qquad a_1 > -\tfrac{13}{80} a_2 \, ,
\eeq
where the second equation coincides with the constraint coming from the stability of the scalar
potential (see \eq{bound45only} in \app{boundedpot}).

\subsubsection{16 only}
\label{16only}
When only the $16_H$ { part of the scalar} potential is considered the symmetry is spontaneously broken to
the standard $SU(5)$ gauge group.
Applying the the stationary \eq{eqstatnu} we find

\begin{align}
&M^2(\overline{5})= 2 \lambda _2 \chi _R^2 \, ,\nn
\\[0.5ex]
&M^2(10)= 0 \, ,
\label{1016only}\\[0.5ex]
&M^2(1)=
\left(
\begin{array}{cc}
 1 & 1 \\
 1 & 1
\end{array}
\right)
\frac{1}{2}\lambda_1\chi_R^2 \, ,\nn
\end{align}
in the ($\psi^{16}_{1}$, $\psi^{16}_{1^\ast}$) basis, { that yields}
\begin{align}
&\text{Det}\ M^2(1)= 0 \, ,
\nn \\[0ex]
&\Tr M^2(1)=\lambda_1\chi_R^2 \, ,
\end{align}
and as expected we count 21 WGB and 10 PGB modes whose mass depends
on $\lambda_2$ only.
The required positivity of the scalar masses leads to
\beq
\label{su5only16cond}
\lambda_2 > 0 \qquad \text{and} \qquad \lambda_1 > 0 \, ,
\eeq
where the second equation coincides with the constraint coming from the stability of the scalar potential
(see \eq{boundonly16} in \app{boundedpot}).

\subsubsection{Mixed 45-16 spectrum ($\chi_R \neq 0$)}
\label{4516spectrum}

In the general case the unbroken symmetry is the SM group. Applying
first the two stationary conditions in \eq{eqstatmu} and \eq{eqstatnu} we find the spectrum below.
The $2\times 2$ matrices are spanned over the ($\psi^{45}$, $\psi^{16}$) basis whereas the $4\times 4$ SM singlet matrix is given in the
($\psi^{45}_{24}$, $\psi^{45}_{1}$, $\psi^{16}_{1}$, $\psi^{16}_{1^\ast}$) basis.

\bwt
\begin{align}
\label{mass1114516}
&M^2(1,1,+1)=
\left(
\begin{array}{cc}
 \beta  \chi _R^2+2 a_2 \omega _Y \left(\omega _R+\omega _Y\right) & \chi _R \left(\tau -3 \beta  \omega _Y\right) \\
 \chi _R \left(\tau -3 \beta  \omega _Y\right) & 2 \omega _R \left(\tau -3 \beta  \omega _Y\right)
\end{array}
\right) \, ,
\nn\\[0ex]
&M^2(\overline{3},1,-\tfrac{2}{3})=
\left(
\begin{array}{cc}
 \beta  \chi _R^2+2 a_2 \omega _R \left(\omega _R+\omega _Y\right) & \chi _R \left(\tau - \beta  (2\omega _R+\omega _Y)\right) \\
 \chi _R \left(\tau - \beta  (2\omega _R+\omega _Y)\right) & 2 \omega _Y \left(\tau - \beta  (2\omega _R+\omega _Y)\right)
\end{array}
\right) \, ,
\\[1ex]
%\end{align}
%\begin{align}
%\label{mass1304516}
&M^2(1,3,0)=
2 a_2 (\omega _Y - \omega _R) (\omega _Y + 2 \omega _R) \, , \nn \\[0ex]
\label{mass8104516}
&M^2(8,1,0)=
2 a_2 (\omega _R - \omega _Y) (\omega _R + 2 \omega _Y) \, , \\[0ex]
&M^2(3,2,-\tfrac{5}{6})=
0 \, ,
\nn\\[1ex]
&M^2(3,2,+\tfrac{1}{6})=
\left(
\begin{array}{cc}
 \beta  \chi _R^2+4 a_2 \omega _R \omega _Y & \chi _R \left(\tau -\beta  (\omega _R+2\omega _Y)\right) \\
 \chi _R \left(\tau  -\beta  (\omega _R+2\omega _Y)\right) & \left(\omega _R+\omega _Y\right) \left(\tau  -\beta  (\omega _R+2\omega _Y)\right)
\end{array}
\right) \, ,
\nn\\[0ex]
\label{mass12m124516}
%\end{align}
%\begin{align}
&M^2(1,2,-\tfrac{1}{2})=
\left(\omega _R+3 \omega _Y\right) \left(\tau -\beta  \omega _R\right)+2 \lambda _2 \chi _R^2
 \, , \\[0ex]
\label{mass3bar1134516}
&M^2(\overline{3},1,+\tfrac{1}{3})=
2 \left(\omega _R+\omega _Y\right) \left(\tau -\beta  \omega _Y\right)+2 \lambda _2 \chi _R^2
 \, \nn.
\end{align}
%\ewt
%\bwt
\begin{multline}
M^2(1,1,0)= \\[1ex]
\left(
\begin{array}{cc}
 \frac{1}{2} \left(3 \beta  \chi _R^2+4 \left(a_2 \omega _R^2+a_2 \omega _Y \omega _R+(48 a_1+7 a_2) \omega _Y^2\right)\right) &
   \sqrt{6} \left(\frac{\beta  \chi _R^2}{2}+2 (16 a_1+3 a_2) \omega _R \omega _Y\right)  \\
 \sqrt{6} \left(\frac{\beta  \chi _R^2}{2}+2 (16 a_1+3 a_2) \omega _R \omega _Y\right) &
 \beta  \chi _R^2+2 \left(4 (8 a_1+a_2) \omega _R^2+a_2 \omega _Y \omega _R+a_2 \omega _Y^2\right)  \\
 -\frac{1}{2} \sqrt{3} \chi _R \left(\tau -2 \beta  \omega _R-(16 \alpha +3 \beta ) \omega _Y\right) &
 \frac{\chi _R \left(-\tau +2 (8 \alpha +\beta ) \omega _R+3 \beta  \omega _Y\right)}{\sqrt{2}}   \\
 -\frac{1}{2} \sqrt{3} \chi _R \left(\tau -2 \beta  \omega _R-(16 \alpha +3 \beta ) \omega _Y\right) &
 \frac{\chi _R \left(-\tau +2 (8 \alpha +\beta ) \omega _R+3 \beta  \omega _Y\right)}{\sqrt{2}}
 \end{array}
\right.  \\[1ex]
 \left.
\begin{array}{cc}
   -\frac{1}{2} \sqrt{3} \chi _R \left(\tau -2 \beta \omega _R-(16 \alpha +3 \beta ) \omega _Y\right) &
   -\frac{1}{2} \sqrt{3} \chi _R \left(\tau -2 \beta  \omega _R-(16 \alpha +3 \beta ) \omega _Y\right) \\
 \frac{\chi _R \left(-\tau +2 (8 \alpha +\beta ) \omega _R+3 \beta  \omega _Y\right)}{\sqrt{2}} &
 \frac{\chi _R \left(-\tau +2 (8 \alpha +\beta ) \omega _R+3 \beta  \omega _Y\right)}{\sqrt{2}} \\
 \frac{1}{2} \lambda _1 \chi _R^2 &
 \frac{1}{2} \lambda _1 \chi _R^2 \\
 \frac{1}{2} \lambda _1 \chi _R^2 &
 \frac{1}{2} \lambda _1 \chi _R^2
\end{array}
\right) \, .
\end{multline}
\ewt

By applying the remaining stationary condition in \eq{eqstat0}
one obtains
\bwt
\begin{align}
&\text{Det}\ M^2(1,1,+1)=0 \, ,
\nn\\[0ex]
%\label{massTr1114516}
&\Tr M^2(1,1,+1)=
\frac{\left(\chi _R^2+4 \omega _R^2\right) \left(\tau -3 \beta  \omega _Y\right)}{2 \omega _R} \, ,
\nn\\[0ex]
&\text{Det}\ M^2(\overline{3},1,-\tfrac{2}{3})=0 \, , \nn\\[0ex]
\label{massTr31m134516}
&\Tr M^2(\overline{3},1,-\tfrac{2}{3})=
\frac{\left(\chi _R^2+4 \omega _Y^2\right) \left(\tau - \beta (2 \omega _R +  \omega _Y) \right)}{2 \omega _Y} \, , \\[0ex]
&\text{Det}\ M^2(3,2,+\tfrac{1}{6})=0 \, , \nn\\[0ex]
%\label{massTr32164516}
&\Tr M^2(3,2,+\tfrac{1}{6})=\beta  \chi _R^2 + 4 a_2 \omega _R \omega _Y+
\left(\omega _R+\omega _Y\right) \left(\tau - \beta  (\omega _R + 2  \omega _Y) \right) \, ,\nn \\[0ex]
%\label{massRank11064516}
&\text{Rank}\ M^2(1,1,0)= 3 \, , \nn\\[0ex]
%\label{massTr11064516}
&\Tr M^2(1,1,0)=2 \left((32 a_1+5 a_2) \omega _R^2
+8 (6 a_1+a_2) \omega _Y^2+2 a_2 \omega _R \omega _Y\right)+\chi _R^2 \left(\tfrac{5}{2}\beta +\lambda _1\right) \, .\nn
\end{align}
\ewt
In \eqs{mass1114516}{massTr31m134516} we recognize the 33 WGB with the quantum numbers of the coset $SO(10)/SM$ algebra.

In using the stationary condition in \eq{eqstat0},
we paid attention not to divide by ($\omega_R + \omega_Y$), since the flipped vacuum $\omega=\omega_R=-\omega_Y$
is an allowed configuration. On the other hand, we can freely { put
 $\omega_R$ and $\omega_Y$ into} the denominators, as
the vacua $\omega_R=0$ and $\omega_Y=0$ are excluded at the tree level.
The coupling $a_2$ in \eq{massTr31m134516} is understood to obey the constraint
\beq
\label{laststatcond}
4a_2(\omega_R+\omega_Y)\omega_R\omega_Y+\beta\chi_R^2(2\omega_R + 3\omega_Y)
-\tau\chi_R^2 = 0 \, .
\eeq

\subsubsection{A trivial 45-16 potential ($a_2=\lambda_2=\beta=\tau=0$)}
\label{4516justnorms}

It is interesting to study the global symmetries of the scalar potential when only the moduli of $45_H$ and $16_H$ appear in the { scalar potential}.
In order to correctly count the corresponding PGB, the $(1,1,0)$ mass matrix in the limit of $a_2=\lambda_2=\beta=\tau=0$ needs to be scrutinized. We find in the ($\psi^{45}_{24}$, $\psi^{45}_{1}$, $\psi^{16}_{1}$, $\psi^{16}_{1^\ast}$) basis,
\bwt
\begin{align}
\label{}
M^2(1,1,0)=&
\left(
\begin{array}{cccc}
 96 a_1 \omega _Y^2 & 32 \sqrt{6} a_1 \omega _R \omega _Y & 8 \sqrt{3} \alpha  \chi _R \omega _Y & 8 \sqrt{3} \alpha  \chi _R \omega _Y \\
 32 \sqrt{6} a_1 \omega _R \omega _Y & 64 a_1 \omega _R^2 & 8 \sqrt{2} \alpha  \chi _R \omega _R & 8 \sqrt{2} \alpha  \chi _R \omega _R \\
 8 \sqrt{3} \alpha  \chi _R \omega _Y & 8 \sqrt{2} \alpha  \chi _R \omega _R & \frac{1}{2} \lambda _1 \chi _R^2 & \frac{1}{2} \lambda _1 \chi _R^2 \\
 8 \sqrt{3} \alpha  \chi _R \omega _Y & 8 \sqrt{2} \alpha  \chi _R \omega _R & \frac{1}{2} \lambda _1 \chi _R^2 & \frac{1}{2} \lambda _1 \chi _R^2
\end{array}
\right)
\, ,
\label{110simplified}
\end{align}
\ewt
with the properties

\begin{align}
&\text{Rank}\ M^2(1,1,0)= 2 \, ,
\nn \\[0ex]
&\Tr M^2(1,1,0)=64 a_1 \omega _R^2+96 a_1 \omega _Y^2+\lambda _1 \chi _R^2 \, .
\end{align}

{ As expected from the discussion in \sect{sect:understanding},  \eqs{mass1114516}{110simplified} in the $a_2=\lambda_2=\beta=\tau=0$ limit exhibit 75 massless
modes out of which 42 are PGB.}

\subsubsection{A trivial 45-16 interaction ($\beta=\tau=0$)}
\label{4516trivialinteraction}

{ In this limit, the interaction part of the potential consists only of} the $\alpha$ term,
which is the product of $45_H$ and  $16_H$ moduli.
Once again, in order to correctly count the massless modes
we specialize the $(1,1,0)$ matrix to the
$\beta=\tau=0$ limit. In the ($\psi^{45}_{24}$, $\psi^{45}_{1}$, $\psi^{16}_{1}$, $\psi^{16}_{1^\ast}$) basis, we find
\bwt
\begin{multline}
M^2(1,1,0)= \\[1ex]
\left(
\begin{array}{cc}
 2 \left(a_2 \omega _R^2+a_2 \omega _Y \omega _R+(48 a_1+7 a_2) \omega _Y^2\right) & 2 \sqrt{6} (16 a_1+3 a_2) \omega _R
   \omega _Y  \\
 2 \sqrt{6} (16 a_1+3 a_2) \omega _R \omega _Y & 2 \left(4 (8 a_1+a_2) \omega _R^2+a_2 \omega _Y \omega _R+a_2 \omega
   _Y^2\right) \\
 8 \sqrt{3} \alpha  \chi _R \omega _Y & 8 \sqrt{2} \alpha  \chi _R \omega _R  \\
 8 \sqrt{3} \alpha  \chi _R \omega _Y & 8 \sqrt{2} \alpha  \chi _R \omega _R
\end{array}
\right.
%\\[0ex]
\left.
\begin{array}{cc}
8 \sqrt{3} \alpha  \chi _R \omega _Y & 8 \sqrt{3} \alpha  \chi _R \omega _Y \\
8 \sqrt{2} \alpha  \chi _R \omega _R & 8 \sqrt{2} \alpha  \chi _R \omega _R \\
 \frac{1}{2} \lambda _1 \chi _R^2 & \frac{1}{2} \lambda _1 \chi _R^2 \\
  \frac{1}{2} \lambda _1 \chi _R^2 & \frac{1}{2} \lambda _1 \chi _R^2
\end{array}
\right)
\, ,
\end{multline}
\ewt
with the properties
\begin{align}
&\text{Rank}\ M^2(1,1,0)= 3  \, , \nn\\[0.5ex]
&\Tr M^2(1,1,0) = 2 \left((32 a_1 + 5 a_2) \omega _R^2 \right.
 +8 (6 a_1+a_2) \omega _Y^2 \nn\\[0.2ex]
&\hspace*{7em} +\left. 2 a_2 \omega _R \omega _Y\right)+\lambda _1 \chi _R^2 \, .
\end{align}

According to the discussion in \sect{sect:understanding}, { upon inspecting}
\eqs{mass1114516}{massTr31m134516} in the $\beta=\tau=0$ limit,
one finds 41 massless scalar modes of which 8 are PGB.

\subsubsection{The 45-16 scalar spectrum for $\chi_R = 0$ }

\label{4516spectrumchi0}

The application of the stationary conditions in \eqs{eqstatmu}{eqstat0}
(for $\chi_R =0$, \eq{eqstatnu} is trivially satisfied) leads to four different spectra according to the four vacua:
standard $5 \, 1_{Z}$,
flipped $5' \, 1_{Z'}$, $3_c\, 2_L\, 2_R\, 1_X$ and $4_C\, 2_L\, 1_R$.
We specialize our discussion to the last three cases.

The mass eigenstates are conveniently labeled according to
the subalgebras of $SO(10)$ left invariant by each vacuum.
{ With the help of \Tables{tab:16decomp}{tab:45decomp}} one can easily recover the decomposition in the SM components.
In the limit $\chi_R=0$ the states $45_H$ and $16_H$ do not mix.
All of the WGB belong to the $45_H$, since for $\chi_R=0$ the $16_H$ preserves $SO(10)$.

Consider first the case: $\omega = \omega_R = -\omega_Y$ (which preserves the flipped $5' \, 1_{Z'}$ group). For the $45_H$ components we obtain:
\begin{align}
&M^2(24,0)= -4 a_2 \omega ^2 \, , \nn\\[0.5ex]
&M^2(10,-4) = 0 \, , \\[0.5ex]
&M^2(1,0) = 2 \left(80 a_1 + 13 a_2\right) \omega ^2 \, .\nn
\end{align}

Analogously, for the $16_H$ components we get:
\begin{align}
&M^2(10,+1) = \tfrac{1}{4} \left(\omega^2 (80 \alpha +\beta )+2 \tau  \omega -2 \nu^2 \right) \, , \nn\\[0.5ex]
&M^2(\bar 5,-3) = \tfrac{1}{4} \left(\omega^2 (80 \alpha +9 \beta )-6 \tau  \omega -2 \nu^2 \right) \, , \\[0.5ex]
&M^2(1,+5) = \tfrac{1}{4} \left(5 \omega^2 (16 \alpha +5 \beta )+10 \tau  \omega -2 \nu^2 \right) \, .\nn
\end{align}
Since the unbroken group is the flipped $5' \, 1_{Z'}$ we recognize, as
expected, 45-25=20 WGB.
When only trivial $45_H$ invariants (moduli) are considered
the global symmetry of the scalar potential is $O(45)$,
broken spontaneously by $\omega$ to $O(44)$.
This leads to 44 GB in the scalar spectrum.
Therefore 44-20=24 PGB are left in the spectrum.
On general grounds, their masses should receive contributions
from all of the explicitly breaking terms $a_2$, $\beta$ and $\tau$.
As it is directly seen from the spectrum, only the
$a_2$ term contributes at the tree level to $M(24,0)$.
By choosing $a_2<0$ one may obtain a consistent minimum of the scalar
potential. Quantum corrections are not relevant in this case.

Consider then the case $\omega_R=0$ and $\omega_Y \neq 0$ which preserves the $3_c\, 2_L\, 2_R\, 1_X$ gauge group.
For the $45_H$ components we obtain:
\begin{align}
&M^2(1,3,1,0)= 2 a_2 \omega _Y^2 \, , \nn\\[0.5ex]
&M^2(1,1,3,0)= 2 a_2 \omega _Y^2 \, , \nn\\[0.5ex]
&M^2(8,1,1,0)= -4 a_2 \omega _Y^2 \, , \nn\\[0.5ex]
&M^2(3,2,2,-\tfrac{1}{3})= 0 \, , \\[0.5ex]
&M^2(\overline{3},1,1,-\tfrac{2}{3})= 0 \, , \nn\\[0.5ex]
&M^2(1,1,1,0)= 2 \left(48 a_1 + 7 a_2\right) \omega _Y^2 \, .\nn
\end{align}

Analogously, for the $16_H$ components we get:
\begin{align}
&M^2(3,2,1,+\tfrac{1}{6})= \tfrac{1}{4} \left( \omega _Y^2(48 \alpha +\beta )
-2 \tau \omega _Y-2 \nu ^2\right) \, , \nn\\[0.5ex]
&M^2(\overline{3},1,2,-\tfrac{1}{6})= \tfrac{1}{4} \left( \omega _Y^2(48 \alpha +\beta )
+2 \tau \omega _Y-2 \nu ^2\right) \, , \nn\\[0.5ex]
&M^2(1,2,1,-\tfrac{1}{2})= \tfrac{1}{4} \left(\omega _Y^2 (48 \alpha +9 \beta )+6 \tau  \omega _Y-2 \nu^2\right)
\, , \nn\\[0.5ex]
&M^2(1,1,2,+\tfrac{1}{2})= \tfrac{1}{4} \left( \omega _Y^2(48 \alpha +9 \beta )-6 \tau  \omega_Y-2 \nu ^2\right) \, .
\end{align}

Worth of a note is the mass degeneracy of
the $(1,3,1,0)$ and $(1,1,3,0)$ multiplets { which is due to the fact that
for $\omega_R=0$ D-parity}  is conserved by even $\omega_Y$ powers.
On the contrary, in the $16_H$ components the D-parity is broken by the $\tau$ term that is linear in $\omega_Y$.

Since the unbroken group is $3_c\, 2_L, 2_R\, 1_X$ there are 45-15=30 WGB, as it appears from the explicit pattern of the scalar spectrum.
When only trivial invariants (moduli terms) of $45_H$ are considered
the global symmetry of the scalar potential is $O(45)$,
broken spontaneously to $O(44)$, thus
leading to 44 GB in the scalar spectrum.
As a consequence 44-30=14 PGB are left in the spectrum.
On general grounds, their masses should receive contributions
from all of the explicitly breaking terms $a_2$, $\beta$ and $\tau$.
As it is directly seen from the spectrum, only the
$a_2$ term contributes at the tree level to the mass of the 14 PGB,
leading unavoidably to a tachyonic spectrum. This feature is naturally lifted at the quantum level.

Let us finally consider the case $\omega_R \neq 0$ and $\omega_Y = 0$ (which preserves the $4_C\, 2_L\, 1_R$ gauge symmetry).
For the $45_H$ components we find:
\begin{align}
&M^2(15,1,0)= 2 a_2 \omega _R^2 \, , \nn\\[0.5ex]
&M^2(1,3,0)= -4 a_2 \omega _R^2 \, , \nn\\[0ex]
&M^2(6,2,+\tfrac{1}{2})= 0 \, , \\[0.5ex]
&M^2(6,2,-\tfrac{1}{2})= 0 \, , \nn\\[0ex]
&M^2(1,1,+1)= 0 \, , \nn\\[0ex]
&M^2(1,1,0)= 8 \left(8 a_1+a_2\right) \omega _R^2 \, .\nn
\end{align}
For the $16_H$ components we obtain:
\begin{align}
&M^2(4,2,0)= 8 \alpha  \omega _R^2 - \tfrac{1}{2}\nu ^2 \, , \nn\\[0.5ex]
&M^2(\overline{4},1,+\tfrac{1}{2})=
\omega _R^2 (8 \alpha +\beta )+\tau  \omega _R - \tfrac{1}{2}\nu ^2 \, , \\[0.5ex]
 &M^2(\overline{4},1,-\tfrac{1}{2})=
\omega _R^2 (8 \alpha +\beta )-\tau  \omega _R - \tfrac{1}{2}\nu ^2 \, .\nn
\end{align}
The unbroken gauge symmetry { in this case corresponds to} $4_C\, 2_L\, 1_R$.
Therefore, one can recognize 45-19=26 WGB in the scalar spectrum.
When only trivial (moduli) $45_H$ invariants are considered
the global symmetry of the scalar potential is $O(45)$,
which is broken spontaneously by $\omega _R$ to $O(44)$.
This leads globally to 44 massless states in the scalar spectrum.
As a consequence, 44-26=18 PGB are left in the $45_H$ spectrum,
that should receive mass contributions
from the explicitly breaking terms $a_2$, $\beta$ and $\tau$.
At the tree level only the $a_2$ term is present,
{ leading again} to a tachyonic spectrum.
This is an accidental tree level feature that is naturally lifted at the quantum level.

%%%%%%%%%%%%%%%%%%%%%%%%%%%%%%%%%%%%%%%%%%%%%%%%%%%%%%%%%%%%%%%%%%%%%%%
\section{One-loop mass spectra}
%%%%%%%%%%%%%%%%%%%%%%%%%%%%%%%%%%%%%%%%%%%%%%%%%%%%%%%%%%%%%%%%%%%%%%%
\label{app:1Lmasses}
{ We have checked explicitly that} the one-loop corrected stationary equation (\ref{eqstat0}) maintains
in the $\chi_R=0$ limit the four tree level solutions, namely,
$\omega_R = \omega_Y$,
$\omega_R = -\omega_Y$, $\omega_R=0$ and $\omega_Y = 0$,
corresponding respectively to the
standard $5\, 1_Z$, flipped $5'\, 1_{Z'}$, $3_c\, 2_L\, 2_R\, 1_X$ and $4_C\, 2_L\, 1_R$ vacua.

In what follows we list, for the last three cases,
the leading one-loop corrections,
arising from the gauge and scalar sectors,
to the critical PGB masses. For all other states
{ the loop corrections provide only sub-leading perturbations of the tree-level masses}, and as such irrelevant
to the present discussion.

\subsection{Gauge contributions to the PGB mass}
\label{gaugePGBoneloop}

Before { focusing to the three relevant vacuum configurations},
it is convenient to write the gauge contribution to the
$(1,3,0)$ and $(8,1,0)$ states
in the general case.

\bwt
\begin{multline}
\Delta M^{2}(1,3,0)=
\frac{g^4 \left(16 \omega _R^2+\omega _Y \omega _R+19 \omega _Y^2\right)}{4 \pi ^2}
+ \frac{3 g^4}{4 \pi ^2\left(\omega _R-\omega _Y\right)}
\left[  2 \left(\omega _R-\omega _Y\right){}^3 \log \left(\frac{g^2 \left(\omega _R-\omega _Y\right){}^2}{\mu ^2}\right) \right. \\[1ex]
+\left(4 \omega
   _R-5 \omega _Y\right) \left(\omega _R+\omega _Y\right){}^2 \log \left(\frac{g^2 \left(\omega _R+\omega _Y\right){}^2}{\mu ^2}\right)
 \left. -4 \omega
   _R^3 \log \left(\frac{4 g^2 \omega _R^2}{\mu ^2}\right)+8 \omega _Y^3 \log \left(\frac{4 g^2 \omega _Y^2}{\mu ^2}\right) \right] \, ,
\end{multline}
\begin{multline}
\Delta M^{2}(8,1,0)=
\frac{g^4\left(13 \omega _R^2+\omega _Y \omega _R+22 \omega _Y^2\right)}{4 \pi ^2}
 +\frac{3 g^4}{8 \pi ^2 \left(\omega _R-\omega _Y\right)}
\left[ \left(\omega _R-\omega _Y\right){}^3 \log \left(\frac{g^2 \left(\omega _R-\omega _Y\right){}^2}{\mu ^2}\right) \right. \\[1ex]
+\left(5 \omega
   _R-7 \omega _Y\right) \left(\omega _R+\omega _Y\right){}^2 \log \left(\frac{g^2 \left(\omega _R+\omega _Y\right){}^2}{\mu ^2}\right)
 \left. +4 \omega
   _Y^2 \left(3 \omega _R+\omega _Y\right) \log \left(\frac{4 g^2 \omega _Y^2}{\mu ^2}\right)-8 \omega _R^3 \log \left(\frac{4 g^2 \omega
   _R^2}{\mu ^2}\right)\right] \, .
\end{multline}
\ewt

One can easily recognize the (tree-level) masses of the
gauge bosons in the log's arguments and cofactors (see \app{gaugespectrum45}). Note that
only the massive states do contribute to the one-loop correction.
(see \sect{sec:1loopspectrum}).

Let's now specialize to the three relevant vacua. { First, for the flipped $5' \, 1_{Z'}$ case $\omega = \omega_R = -\omega_Y$ one has}:
\begin{multline}
\Delta M^{2}(24,0)
= \frac{17 g^4 \omega ^2}{2 \pi ^2}
+\frac{3 g^4 \omega ^2 }{2 \pi ^2}\log \left(\frac{4 g^2 \omega ^2}{\mu ^2}\right) \, .
\end{multline}
Similarly, for $\omega_R=0$ and $\omega_Y \neq 0$\quad ($3_c\, 2_L\, 2_R\, 1_X$):
\begin{align}
&\Delta M^{2}(1,3,1,0)  = \Delta M^{2}(1,1,3,0) =
\frac{19 g^4 \omega _Y^2}{4 \pi ^2} \\[0.5ex]
&\qquad +\frac{21 g^4 \omega _Y^2}{4 \pi ^2} \log \left(\frac{g^2 \omega _Y^2}{\mu ^2}\right)
-\frac{24 g^4 \omega _Y^2}{4 \pi ^2} \log \left(\frac{4 g^2 \omega _Y^2}{\mu^2}\right) \, ,\nn
\end{align}
\beq
\Delta M^{2}(8,1,1,0) =
\frac{11 g^4 \omega _Y^2}{2 \pi ^2}
+\frac{3 g^4 \omega _Y^2}{2 \pi ^2} \log \left(\frac{g^2 \omega _Y^2}{4 \mu ^2}\right)\, .
\eeq

Finally, for $\omega_R\neq 0$ and $\omega_Y = 0$\quad ($4_C\, 2_L\, 1_R$):
\beq
\Delta M^{2}(1,3,0)  =
\frac{4 g^4 \omega _R^2}{\pi ^2}
+\frac{3 g^4 \omega _R^2}{2 \pi ^2} \log \left(\frac{g^2 \omega _R^2}{16 \mu ^2}\right)\, ,
\eeq
\begin{multline}
\Delta M^{2}(15,1,0) =
\frac{13 g^4 \omega _R^2}{4 \pi ^2}
+\frac{9 g^4 \omega _R^2}{4 \pi ^2} \log \left(\frac{g^2 \omega _R^2}{\mu ^2}\right) \\[0.5ex]
-\frac{12 g^4 \omega _R^2}{4 \pi ^2}  \log \left(\frac{4 g^2 \omega _R^2}{\mu ^2}\right)\, .
\end{multline}

\subsection{Scalar contributions to the PGB mass}
\label{scalarPGBoneloop}

Since the general formula { for the SM vacuum configuration is quite involved,
we give directly the corrections to the PGB masses on the three vacua of our interest}.
We consider first the case $\omega = \omega_R = -\omega_Y$ (flipped $5' \, 1_{Z'}$):
\bwt
\begin{align}
\Delta M^{2}&(24,0) =
\frac{\tau ^2 + 5 \beta ^2 \omega ^2}{4 \pi ^2} \\[0ex]
&+ \frac{1}{128 \pi ^2 \omega }\left[
 (-5 \beta  \omega -\tau ) (5 \omega  (16 \alpha  \omega +5 \beta  \omega +2 \tau )-2 \nu^2)
\log \left(\frac{5 \omega^2 (16 \alpha +5 \beta )+10 \tau  \omega -2 \nu^2}{4 \mu ^2}\right) \right. \nn \\[0ex]
& + \left(\omega  \left(3 \tau  \omega  (80 \alpha +3 \beta )+\beta  \omega ^2 (27 \beta -400 \alpha )-10 \tau ^2\right)+\nu^2 (10 \beta  \omega
-6 \tau)\right) \log \left(\frac{\omega^2 (80 \alpha +9 \beta )-6 \tau  \omega -2 \nu^2}{4 \mu ^2}\right) \nn \\[0ex]
& \left. + 2 \left(\omega  \left(\tau  (33 \beta  \omega -80 \alpha  \omega )+\beta \omega ^2 (400 \alpha +17 \beta )+10 \tau ^2\right)+2 \nu^2 (\tau -5 \beta \omega )\right) \log \left(\frac{\omega^2 (80 \alpha +\beta )+2 \tau  \omega -2 \nu^2 }{4 \mu ^2}\right) \right].\nn
\end{align}
\ewt

For $\omega_R=0$ and $\omega_Y \neq 0$\quad  ($3_c\, 2_L\, 2_R\, 1_X$),
we find:

\bwt
\begin{align}
\Delta M^{2}&(1,3,1,0)  = \Delta M^{2}(1,1,3,0)  =
\frac{\tau ^2 + 2 \beta ^2 \omega _Y^2}{4 \pi ^2}  \\[0ex]
&+ \frac{1}{64 \pi ^2 \omega_Y }\left[
-\left(\tau -3 \beta  \omega _Y\right) \left(-3 \omega _Y^2 (16 \alpha +3 \beta )+6 \tau  \omega _Y+2 \nu^2 \right) \log \left(\frac{ \omega _Y^2(48 \alpha +9 \beta )-6 \tau
   \omega _Y -2 \nu ^2 }{4 \mu ^2}\right) \right.   \nn \\[0ex]
&-\left(\beta  \omega _Y+\tau \right) \left(\omega _Y^2 (48 \alpha +\beta )+2 \tau  \omega _Y-2 \nu^2 \right) \log \left(\frac{\omega _Y^2 (48
   \alpha +\beta )+2 \tau  \omega _Y-2 \nu^2 }{4 \mu ^2}\right)
      \nn \\[0ex]
&+ \left(3 \tau  \omega _Y^2 (16 \alpha -11 \beta )+\beta  \omega _Y^3 (240 \alpha +17 \beta )+2 \omega _Y \left(5 \tau ^2-5 \beta \nu^2 \right)-2
    \nu^2  \tau \right) \log \left(\frac{\omega _Y^2 (48 \alpha +\beta )-2 \tau  \omega _Y-2  \nu^2 }{4 \mu ^2}\right)
    \nn \\[0ex]
&\left. +\left(\omega _Y^2 (9 \beta  \tau -48 \alpha  \tau )+3 \beta  \omega _Y^3 (9 \beta -16 \alpha )+2 \omega _Y \left(\beta   \nu^2 -\tau ^2\right)+2
    \nu^2  \tau \right) \log \left(\frac{\omega _Y^2 (48 \alpha +9 \beta )+6 \tau  \omega _Y-2  \nu^2 }{4 \mu ^2}\right) \right]\nn\,,
\end{align}
\begin{align}
\Delta &M^{2}(8,1,1,0) =
\frac{\tau ^2 + 3 \beta ^2 \omega _Y^2}{4 \pi ^2}  \\[0ex]
&+ \frac{1}{64 \pi ^2 \omega_Y } \left[
 - \left(\tau -3 \beta  \omega _Y\right) \left(-3 \omega _Y^2 (16 \alpha +3 \beta )+6 \tau  \omega _Y+2  \nu^2 \right) \log \left(\frac{\omega _Y^2 (48
   \alpha +9 \beta )-6 \tau  \omega _Y-2 \nu^2 }{4 \mu ^2}\right) \right. \nn \\[0ex]
& + \left(\omega _Y^2 (21 \beta  \tau -48 \alpha  \tau )+\beta  \omega _Y^3 (144 \alpha +11 \beta )+\omega _Y \left(6 \tau ^2-6 \beta  \nu^2 \right)+2
    \nu^2  \tau \right) \log \left(\frac{\omega _Y^2 (48 \alpha +\beta )+2 \tau  \omega _Y-2  \nu^2 }{4 \mu ^2}\right) \nn \\[0ex]
& - \left(3 \beta  \omega _Y+\tau \right) \left(\omega _Y^2 (48 \alpha +9 \beta )+6 \tau  \omega _Y-2  \nu^2 \right) \log \left(\frac{\omega _Y^2 (48 \alpha
   +9 \beta )+6 \tau  \omega _Y-2  \nu^2 }{4 \mu ^2}\right) \nn \\[0ex]
&\left. + \left(3 \tau  \omega _Y^2 (16 \alpha -7 \beta )+\beta  \omega _Y^3 (144 \alpha +11 \beta )+\omega _Y \left(6 \tau ^2-6 \beta   \nu^2 \right)-2  \nu^2  \tau \right) \log \left(\frac{\omega _Y^2 (48 \alpha +\beta )-2 \tau  \omega _Y-2  \nu^2 }{4 \mu ^2}\right) \right]\nn.
\end{align}
\ewt

Finally, for $\omega_R\neq 0$ and $\omega_Y = 0$\quad ($4_C\, 2_L\, 1_R$), we have:
\bwt
\begin{align}
\Delta M^{2}&(1,3,0)  =
\frac{\tau ^2 + 2 \beta ^2 \omega _R^2}{4 \pi ^2} + \frac{1}{64 \pi ^2 \omega_R } \left[
16 \omega _R \left(16 \alpha  \beta  \omega _R^2-\beta   \nu^2 +\tau ^2\right) \log \left(\frac{8 \alpha  \omega _R^2-\frac{\nu ^2}{2}}{\mu ^2}\right) \right.  \nn \\[0ex]
&-4 \left(\tau -2 \beta  \omega _R\right) \left(-2 \omega _R^2 (8 \alpha +\beta )+2 \tau  \omega _R+ \nu^2 \right)
\log \left(\frac{ \omega _R^2 (8 \alpha +\beta )-\tau  \omega _R -\frac{\nu ^2}{2}}{ \mu ^2}\right)   \nn \\[0ex]
&\left. -4 \left(2 \beta  \omega _R+\tau \right) \left(2 \omega _R^2 (8 \alpha +\beta )+2 \tau  \omega _R- \nu^2 \right)
\log \left(\frac{ \omega _R^2 (8 \alpha +\beta )+\tau  \omega _R -\frac{\nu ^2}{2} }{4 \mu ^2}\right) \right]\,,
\\
%\end{align}
%\begin{align}
\Delta M^{2}&(15,1,0) =
\frac{\tau ^2 + \beta ^2 \omega _R^2}{4 \pi ^2} \nn \\[0ex]
&+ \frac{1}{64 \pi ^2 \omega_R } \left[
 8 \omega _R \left(16 \alpha  \beta  \omega _R^2-\beta  \nu^2 +\tau ^2\right) \log \left(\frac{8 \alpha  \omega _R^2-\frac{\nu ^2}{2}}{\mu ^2}\right) \right. \nn \\[0ex]
& -4 \left(2 \beta  \omega _R^3 (8 \alpha -\beta )-16 \alpha  \tau  \omega _R^2+\omega _R \left(\tau ^2-\beta  \nu^2 \right)+ \nu^2  \tau \right)
   \log \left(\frac{ \omega _R^2 (8 \alpha +\beta )-\tau  \omega _R -\frac{\nu ^2}{2}}{\mu ^2}\right) \nn \\[0ex]
& \left. + 4 \left(2 \beta  \omega _R^3 (\beta -8 \alpha )-16 \alpha  \tau  \omega _R^2+\omega _R \left(\beta  \nu^2 -\tau ^2\right)+ \nu^2  \tau \right)
   \log \left(\frac{ \omega _R^2 (8 \alpha +\beta )+\tau  \omega _R -\frac{\nu ^2}{2}}{\mu ^2}\right) \right]\,.
\end{align}
\ewt

{ Also in these formulae} we recognize  the (tree level)
mass eigenvalues of the $16_H$ states contributing to the one-loop
effective potential
(see \app{4516spectrumchi0}).

Notice that the { singlets with respect to each vacuum},
namely $(1,0)$, $(1,1,1,0)$ and $(1,1,0)$, for the flipped $5' \, 1_{Z'}$,
$3_c\, 2_L\, 2_R\, 1_X$ and $4_C\, 2_L\, 1_R$ vacua respectively,
receive a tree level contribution { from both $a_1$ as well as} $a_2$
(see \app{4516spectrumchi0}).
The $a_1$ term leads the tree level mass and
radiative corrections can be neglected.

One may verify that in the limit of vanishing VEVs
the one-loop masses vanish identically on each of the three vacua, as it should be. This is a non trivial check of the calculation
of the scalar induced corrections.

%\clearpage


\vfill
\begin{thebibliography}{99}



%%%%%%%%%%-GUT-%%%%%%%%%%%%%%%%%%%%

\bibitem{GUT}
  H.~Georgi, in {\it Particles and Fields}, edited by C.~E.~Carlson
  (AIP, New York, 1975);

%\bibitem{Fritzsch:1974nn}
  H.~Fritzsch and P.~Minkowski,
  {\it Unified Interactions Of Leptons And Hadrons,}
  Annals Phys.\  {\bf 93} (1975) 193.
  %%CITATION = APNYA,93,193;%%


%%%%%%%%%%-SEESAW I-%%%%%%%%%%%%%%%%%%%%%%%%%%%%%%%%%%%%%%%%%%%%%%%%%%%%%%%%%%%%%%%

\bibitem{seesawI}

%\bibitem{Minkowski:1977sc}
  P.~Minkowski,
  {\it Mu $\to$ E Gamma At A Rate Of One Out Of 1-Billion Muon Decays?,}
  Phys.\ Lett.\ B {\bf 67}, 421 (1977);
  %%CITATION = PHLTA,B67,421;%%

%\bibitem{Gell-Mann:1980vs}
  M.~Gell-Mann, P.~Ramond and R.~Slansky,
  {\it Complex Spinors And Unified Theories,}
  In {\it Supergravity}, P.~van Nieuwenhuizen and D.Z.~Freedman (eds.),
  North Holland Publ.\ Co., 1979, p.\ 315;
%Published in Stony Brook Wkshp.1979:0315 (QC178:S8:1979)

%\bibitem{Yanagida:1979as}
  T.~Yanagida,
  {\it Horizontal Gauge Symmetry And Masses Of Neutrinos,}
  In Proc.\ {\it Workshop on the Baryon Number of the Universe
  and Unified Theories}, O.~Sawada and A.~Sugamoto (eds.), Tsukuba, Japan,
  13--14 Feb.\ 1979, p.\ 95;

%\bibitem{Glashow:1979nm}
  S.L.~Glashow,
  {\it The Future Of Elementary Particle Physics,} HUTP-79-A059
  In Proc.\ Cargese 1979 {\it Quarks and Leptons}, p.\ 687;

%\bibitem{Mohapatra:1979ia}
  R.N.~Mohapatra and G.~Senjanovi\'c,
  {\it Neutrino Mass And Spontaneous Parity Nonconservation,}
  Phys.\ Rev.\ Lett.\  {\bf 44}, 912 (1980).
  %%CITATION = PRLTA,44,912;%%


%%%%%%%%%%-SEESAW II-%%%%%%%%%%%%%%%%%%%%%%%%%%%%%%%%%%%%%%%%%%%%%%%%%%%%%%%%%%%%%%%

\bibitem{seesawII}

%\bibitem{Magg:1980ut}
  M.~Magg and C.~Wetterich,
  {\it Neutrino Mass Problem And Gauge Hierarchy,}
  Phys.\ Lett.\ B {\bf 94}, 61 (1980);
  %%CITATION = PHLTA,B94,61;%%

%\bibitem{Schechter:1980gr}
  J.~Schechter and J.~W.~F.~Valle,
  {\it Neutrino Masses In SU(2) X U(1) Theories,}
  Phys.\ Rev.\  D {\bf 22}, 2227 (1980).
  %%CITATION = PHRVA,D22,2227;%%

%\bibitem{Lazarides:1980nt}
  G.~Lazarides, Q.~Shafi and C.~Wetterich,
  {\it Proton Lifetime And Fermion Masses In An SO(10) Model,}
  Nucl.\ Phys.\ B {\bf 181}, 287 (1981);
  %%CITATION = NUPHA,B181,287;%%

%\bibitem{Mohapatra:1980yp}
  R.~N.~Mohapatra and G.~Senjanovi\'c,
  {\it Neutrino Masses And Mixings In Gauge Models With Spontaneous Parity
  Violation,}
  Phys.\ Rev.\ D {\bf 23}, 165 (1981).
  %%CITATION = PHRVA,D23,165;%%

%bibitem{Schechter:1981cv}
%  J.~Schechter and J.~W.~F.~Valle,
%  {\it Neutrino Decay And Spontaneous Violation Of Lepton Number,}
%  Phys.\ Rev.\  D {\bf 25}, 774 (1982).
  %%CITATION = PHRVA,D25,774;%%


%%%%%%%%%%%%%%%%%%%%NEUTRINOS %%%%%%%%%%%%%%%%%%%%%

\bibitem{neutrinodata}
%\bibitem{Strumia:2006db}
  A.~Strumia and F.~Vissani,
  {\it Neutrino masses and mixings and...,}
  arXiv:hep-ph/0606054;

%\cite{Schwetz:2008er}
  T.~Schwetz, M.~A.~Tortola and J.~W.~F.~Valle,
  {\it Three-flavour neutrino oscillation update,}
  New J.\ Phys.\  {\bf 10} (2008) 113011
  [arXiv:0808.2016 [hep-ph]].
  %%CITATION = NJOPF,10,113011;%%


%%%%%%%%%%-PATI-SALAM-%%%%%%%%%%%%%%%%%%%%%%%%%%%%%%%%%%%%%%%%%%%%%%%%%%%%%%%%%%%%%%%

%bibitem{Pati:1974yy}
%  J.~C.~Pati and A.~Salam,
%  {\it Lepton Number As The Fourth Color,}
%  Phys.\ Rev.\ D {\bf 10} (1974) 275.
  %%CITATION = PHRVA,D10,275;%%


%%%%%%%%%%%%%%%% Minimal SUSY SU(5) orig %%%%%%%%%%%%%%%%%%%%%%%%%%%%%%%%

%bibitem{Dimopoulos:1981zb}
%  S.~Dimopoulos and H.~Georgi,
%  {\it Softly Broken Supersymmetry And SU(5),}
% Nucl.\ Phys.\  B {\bf 193}, 150 (1981).
  %%CITATION = NUPHA,B193,150;%%


%%%%%%%%%%%%%%%% Minimal SUSY SO(10) orig %%%%%%%%%%%%%%%%%%%%%%%%%%%%%%%

\bibitem{MSSO10orig}

%\bibit{Clark:ai}
  T.~E.~Clark, T.~K.~Kuo and N.~Nakagawa,
  {\it A SO(10) Supersymmetric Grand Unified Theory,}
  Phys.\ Lett.\ B {\bf 115} (1982) 26.
  %%CITATION = PHLTA,B115,26;%%

%\bibitem{Aulakh:1982sw}
  C.~S.~Aulakh and R.~N.~Mohapatra,
  {\it Implications Of Supersymmetric SO(10) Grand Unification,}
  Phys.\ Rev.\ D {\bf 28}, 217 (1983).
  %%CITATION = PHRVA,D28,217;%%


%%%%%%%% Minimal SUSY SO(10) recent %%%%%%%%%%%%%%%%%%%%%%%%%%%%%%%%%%%%%%%%%%%%%%%%%%%%%%%%%

\bibitem{MSSO10recent}

%bibitem{Aulakh:2000sn}
%  C.~S.~Aulakh, B.~Bajc, A.~Melfo, A.~Rasin and G.~Senjanovic,
%  {\it SO(10) theory of R-parity and neutrino mass,}
%  Nucl.\ Phys.\ B {\bf 597} (2001) 89
  %[hep-ph/0004031].
  %%CITATION = HEP-PH 0004031;%%

%\bibitem{Fukuyama:2002ch}
  T.~Fukuyama and N.~Okada,
  {\it Neutrino oscillation data versus minimal supersymmetric SO(10) model,}
  JHEP {\bf 0211} (2002) 011
  %[hep-ph/0205066].
  %%CITATION = HEP-PH 0205066;%%

%\bibitem{Bajc:2002iw}
  B.~Bajc, G.~Senjanovic and F.~Vissani,
  {\it b - tau unification and large atmospheric mixing:
  A case for non-canonical see-saw,}
  Phys.\ Rev.\ Lett.\  {\bf 90} (2003) 051802
  %[hep-ph/0210207].
  %%CITATION = HEP-PH 0210207;%%

%\bibitem{Goh:2003sy}
  H.~S.~Goh, R.~N.~Mohapatra and S.~P.~Ng,
  {\it Minimal SUSY SO(10), b tau unification and large neutrino mixings,}
  Phys.\ Lett.\ B {\bf 570}, 215 (2003)
  %[hep-ph/0303055].
  %%CITATION = HEP-PH 0303055;%%

%\bibitem{Fukuyama:2003hn}
  T.~Fukuyama, T.~Kikuchi and N.~Okada,
  {\it Lepton flavor violating processes and muon g-2 in minimal
  supersymmetric SO(10) model,}
  Phys.\ Rev.\ D {\bf 68} (2003) 033012
  %[hep-ph/0304190];
  %%CITATION = HEP-PH 0304190;%%

%\bibitem{Aulakh:2003kg}
  C.~S.~Aulakh, B.~Bajc, A.~Melfo, G.~Senjanovic and F.~Vissani,
  {\it The minimal supersymmetric grand unified theory,}
  Phys.\ Lett.\ B {\bf 588} (2004) 196
  %[hep-ph/0306242].
  %%CITATION = HEP-PH 0306242;%%

%\bibitem{Goh:2003hf}
  H.~S.~Goh, R.~N.~Mohapatra and S.~P.~Ng,
  {\it Minimal SUSY SO(10) model and predictions for neutrino mixings and leptonic CP violation,}
  Phys.\ Rev.\ D {\bf 68}, 115008 (2003)
  %[hep-ph/0308197].
  %%CITATION = HEP-PH 0308197;%%

%\bibitem{Fukuyama:2004xs}
  T.~Fukuyama, A.~Ilakovac, T.~Kikuchi, S.~Meljanac and N.~Okada,
  {\it General formulation for proton decay rate in minimal supersymmetric SO(10) GUT,}
  Eur.\ Phys.\ J.\  C {\bf 42}, 191 (2005)
  %[arXiv:hep-ph/0401213].
  %%CITATION = EPHJA,C42,191;%%

%\bibitem{Bajc:2004fj}
  B.~Bajc, G.~Senjanovic and F.~Vissani,
  {\it Probing the nature of the seesaw in renormalizable SO(10),}
  Phys.\ Rev.\ D {\bf 70} (2004) 093002
  %[hep-ph/0402140].
  %%CITATION = HEP-PH 0402140;%%

%\bibitem{Dutta:2004wv}
  B.~Dutta, Y.~Mimura and R.~N.~Mohapatra,
  {\it CKM CP violation in a minimal SO(10) model for neutrinos and its
  implications,}
  Phys.\ Rev.\ D {\bf 69} (2004) 115014
  %[hep-ph/0402113].
  %%CITATION = HEP-PH 0402113;%%

%\bibitem{Goh:2004fy}
  H.~S.~Goh, R.~N.~Mohapatra and S.~Nasri,
  {\it SO(10) symmetry breaking and type II seesaw,}
  Phys.\ Rev.\ D {\bf 70}, 075022 (2004)
  %[arXiv:hep-ph/0408139].
  %%CITATION = HEP-PH 0408139;%%

%\bibitem{Aulakh:2004hm}
  C.~S.~Aulakh and A.~Girdhar,
  {\it SO(10) MSGUT: spectra, couplings and thresholds effects,}
  Nucl.\ Phys.\ B {\bf 711}, 275 (2005)
  %[hep-ph/0405074].
  %%CITATION = HEP-PH 0405074;%%

%\bibitem{Fukuyama:2004ti}
  T.~Fukuyama, A.~Ilakovac, T.~Kikuchi, S.~Meljanac and N.~Okada,
  {\it Higgs masses in the minimal SUSY SO(10) GUT,}
  Phys.\ Rev.\ D {\bf 72} (2005) 051701
  %[hep-ph/0412348].
  %%CITATION = HEP-PH 0412348;%%

%\bibitem{Aulakh:2005ic}
  C.~S.~Aulakh,
  {\it Consistency of the minimal supersymmetric GUT spectra,}
  Phys.\ Rev.\ D {\bf 72}, 051702 (2005).
  %%CITATION = PHRVA,D72,051702;%%

%\bibitem{Bertolini:2005qb}
  S.~Bertolini and M.~Malinsky,
  {\it On CP violation in a minimal renormalizable SUSY SO(10) model and beyond},
  Phys.\ Rev.\ D {\bf 72}, 055021 (2005)
  %[hep-ph/0504241].
  %%CITATION = HEP-PH 0504241;%%

%\bibitem{Babu:2005ia}
  K.~S.~Babu and C.~Macesanu,
  {\it Neutrino masses and mixings in a minimal SO(10) model,}
  Phys.\ Rev.\ D {\bf 72} (2005) 115003
  %[hep-ph/0505200].
  %%CITATION = HEP-PH 0505200;%%

%\bibitem{Dutta:2005ni}
  B.~Dutta, Y.~Mimura and R.~N.~Mohapatra,
  {\it Neutrino mixing predictions of a minimal SO(10) model with
  suppressed proton decay,}
  Phys.\ Rev.\ D {\bf 72}, 075009 (2005)
  %[hep-ph/0507319].
  %%CITATION = HEP-PH 0507319;%%

%\bibitem{Bajc:2005qe}
  B.~Bajc, A.~Melfo, G.~Senjanovic and F.~Vissani,
  {\it Fermion mass relations and the structure of the light Higgs in a
  supersymmetric SO(10) theory,}
  Phys.\ Lett.\ B, {\bf 634} (2006) 272
  %[hep-ph/0511352].
  %%CITATION = HEP-PH 0511352;%%

\bibitem{Bajc:2004xe}
  B.~Bajc, A.~Melfo, G.~Senjanovic and F.~Vissani,
  {\it The minimal supersymmetric grand unified theory. I: Symmetry breaking  and the particle spectrum},
  Phys.\ Rev.\ D {\bf 70}, 035007 (2004)
  %[hep-ph/0402122].
  %%CITATION = HEP-PH 0402122;%%


%%%%%%%%%%%%%%%% Minimal SUSY SU(5) not working %%%%%%%%%%%%%%%%%%%%%%%%%%%%%%%%%%%%

%bibitem{Murayama:2001ur}
%  H.~Murayama and A.~Pierce,
%  {\it Not even decoupling can save minimal supersymmetric SU(5),}
%  Phys.\ Rev.\  D {\bf 65}, 055009 (2002)
%  [arXiv:hep-ph/0108104].
  %%CITATION = PHRVA,D65,055009;%%


%%%%%%%%%%%%%%%% Minimal SUSY SO(10) not working %%%%%%%%%%%%%%%%%%%%%%%%%%%%%%%%%%%%

\bibitem{MSSO10notworking}

%\bibitem{Aulakh:2005mw}
  C.~S.~Aulakh and S.~K.~Garg,
  {\it MSGUT: From bloom to doom,}
  Nucl.\ Phys.\  B {\bf 757}, 47 (2006)
  [arXiv:hep-ph/0512224].
  %%CITATION = NUPHA,B757,47;%%

%\bibitem{Bertolini:2006pe}
  S.~Bertolini, T.~Schwetz and M.~Malinsky,
  {\it Fermion masses and mixings in SO(10) models and the neutrino challenge  to
  SUSY GUTs,}
  Phys.\ Rev.\  D {\bf 73}, 115012 (2006)
  [arXiv:hep-ph/0605006].
  %%CITATION = PHRVA,D73,115012;%%


%%%%%%%%%%%%%%%% Minimal SUSY SU(5) with NR op's  %%%%%%%%%%%%%%%%%%%%%%%%%%%%%%%%%%%%

%bibitem{Bajc:2002pg}
%  B.~Bajc, P.~Fileviez Perez and G.~Senjanovic,
%  {\it Minimal supersymmetric SU(5) theory and proton decay: Where do we stand?,}
%  arXiv:hep-ph/0210374.
  %%CITATION = HEP-PH/0210374;%%


%%%%%%%%%%%%%%%% MSGUT with 120  %%%%%%%%%%%%%%%%%%%%%%%%%%%%%%%%%%%%

\bibitem{MSGUT120}

%\bibitem{Bertolini:2004eq}
  S.~Bertolini, M.~Frigerio and M.~Malinsky,
  {\it Fermion masses in SUSY SO(10) with type II seesaw: A non-minimal
  predictive scenario,}
  Phys.\ Rev.\  D {\bf 70}, 095002 (2004)
  [arXiv:hep-ph/0406117].
  %%CITATION = PHRVA,D70,095002;%%

%bibitem{Aulakh:2006vi}
%  C.~S.~Aulakh,
%  {\it Fermion mass hierarchy in the Nu MSGUT. I: The real core,}
%  arXiv:hep-ph/0602132.
  %%CITATION = HEP-PH/0602132;%%

%\bibitem{Grimus:2006bb}
  W.~Grimus and H.~Kuhbock,
  {\it Fermion masses and mixings in a renormalizable SO(10) x $Z_2$ GUT,}
  Phys.\ Lett.\  B {\bf 643}, 182 (2006)
  [arXiv:hep-ph/0607197].
  %%CITATION = PHLTA,B643,182;%%

%\bibitem{Aulakh:2007ir}
  C.~S.~Aulakh,
  {\it Pinning down the New Minimal Supersymmetric GUT,}
  Phys.\ Lett.\  B {\bf 661}, 196 (2008)
  %[arXiv:hep-ph/0710.3945].
  %%CITATION = PHLTA,B661,196;%%

%%%%%%%%%%%%%%%% split-SUSY SO(10) %%%%%%%%%%%%%%%%%%%%%%%%%%%%%%%%%%%%

\bibitem{Bajc:2008dc}
  B.~Bajc, I.~Dorsner and M.~Nemevsek,
  {\it Minimal SO(10) splits supersymmetry,}
  JHEP {\bf 0811}, 007 (2008)
  [arXiv:0809.1069 [hep-ph]].
  %%CITATION = JHEPA,0811,007;%%


%%%%%%%%%%%% SUSY BREAKING at M_G ? %%%%%%%%%%%%%%%%%%%%%%%%%%%%%

\bibitem{Hall:2009nd}
  L.~J.~Hall and Y.~Nomura,
  {\it A Finely-Predicted Higgs Boson Mass from A Finely-Tuned Weak Scale,}
  arXiv:0910.2235 [hep-ph].
  %%CITATION = ARXIV:0910.2235;%%


%%%%%%%% NON-SUSY SO10 RGE OLD %%%%%%%%%%%%%%%%%%%%%%%%%%%%%%%%%%%%%%%%%%%%%%%%%%%%%%%%%%%

\bibitem{Gipson:1984aj}
  J.~M.~Gipson and R.~E.~Marshak,
  {\it Intermediate Mass Scales In The New SO(10) Grand Unification In The One Loop Approximation,}
  Phys.\ Rev.\  D {\bf 31}, 1705 (1985).
  %%CITATION = PHRVA,D31,1705;%%

%\bibitem{Chang:1984qr}
  D.~Chang, R.~N.~Mohapatra, J.~Gipson, R.~E.~Marshak and M.~K.~Parida,
  {\it Experimental Tests Of New SO(10) Grand Unification,}
  Phys.\ Rev.\  D {\bf 31}, 1718 (1985).
  %%CITATION = PHRVA,D31,1718;%%

\bibitem{Deshpande:1992au}
  N.~G.~Deshpande, E.~Keith and P.~B.~Pal,
  {\it Implications Of Lep Results For SO(10) Grand Unification,}
  Phys.\ Rev.\  D {\bf 46}, 2261 (1993).
  %%CITATION = PHRVA,D46,2261;%%

%\bibitem{Deshpande:1992em}
  N.~G.~Deshpande, E.~Keith and P.~B.~Pal,
  {\it Implications of LEP results for SO(10) grand unification with two intermediate stages,}
  Phys.\ Rev.\  D {\bf 47}, 2892 (1993)
  %[arXiv:hep-ph/9211232].
  %%CITATION = PHRVA,D47,2892;%%


%%%%%%%%%%%%%%%% NON-SUSY SO(10) RECENT %%%%%%%%%%%%%%%%%%%%%%%%%%%%%%%%%%%%

\bibitem{Bajc:2005zf}
  B.~Bajc, A.~Melfo, G.~Senjanovic and F.~Vissani,
  {\it Yukawa sector in nonsupersymmetric renormalizable SO(10),}
  Phys.\ Rev.\  D {\bf 73}, 055001 (2006)
  %[arXiv:hep-ph/0510139].
  %%CITATION = PHRVA,D73,055001;%%

\bibitem{Bertolini:2009qj}
  S.~Bertolini, L.~Di Luzio and M.~Malinsky,
  {\it Intermediate mass scales in the nonsupersymmetric SO(10) grand
  unification: a reappraisal,}
  Phys.\ Rev.\  D {\bf 80}, 015013 (2009)
  [arXiv:0903.4049 [hep-ph]].
  %%CITATION = PHRVA,D80,015013;%%


%%%%%%%% SO(10) with 144_H %%%%%%%%%%%%%%%%%%%%%%%%%%%%%%%%%%%%%%%%%%%%%%%%%%%%%%%%%

\bibitem{Babu:2005gx}
  K.~S.~Babu, I.~Gogoladze, P.~Nath and R.~M.~Syed,
  {\it A unified framework for symmetry breaking in SO(10),}
  Phys.\ Rev.\  D {\bf 72}, 095011 (2005)
  [arXiv:hep-ph/0506312].
  %%CITATION = PHRVA,D72,095011;%%

%\bibitem{Babu:2006rp}
  K.~S.~Babu, I.~Gogoladze, P.~Nath and R.~M.~Syed,
  {\it Fermion mass generation in SO(10) with a unified Higgs sector,}
  Phys.\ Rev.\  D {\bf 74}, 075004 (2006)
  [arXiv:hep-ph/0607244].
  %%CITATION = PHRVA,D74,075004;%%

\bibitem{Nath:2009nf}
  P.~Nath and R.~M.~Syed,
  {\it Yukawa Couplings and Quark and Lepton Masses in an SO(10) Model with a
  Unified Higgs Sector,}
  arXiv:0909.2380 [hep-ph].
  %%CITATION = ARXIV:0909.2380;%%

%\bibitem{Wu:2009jb}
  Y.~Wu and D.~X.~Zhang,
  {\it Proton Decay and Fermion Masses in Supersymmetric SO(10) Model with Unified
  Higgs Sector,}
  Phys.\ Rev.\  D {\bf 80}, 035022 (2009)
  [arXiv:0909.1179 [hep-ph]].
  %%CITATION = PHRVA,D80,035022;%%


%%%%%%%% WHY 45 + 54 in SUSY %%%%%%%%%%%%%%%%%%%%%%%%%%%%%%%%%%%%%%%%%%%%%%%%%%%%%%%%%

\bibitem{Aulakh:2000sn}
  C.~S.~Aulakh, B.~Bajc, A.~Melfo, A.~Rasin and G.~Senjanovic,
  {\it SO(10) theory of R-parity and neutrino mass,}
  Nucl.\ Phys.\  B {\bf 597}, 89 (2001)
  [arXiv:hep-ph/0004031].
  %%CITATION = NUPHA,B597,89;%%


%%%%%%%% MSO10 (non SUSY) %%%%%%%%%%%%%%%%%%%%%%%%%%%%%%%%%%%%%%%%%%%%%%%%%%%%%%%%%

\bibitem{Yasue}

%\bibitem{Yasue:1980fy}
  M.~Yasue,
  {\it Symmetry Breaking Of SO(10) And Constraints On Higgs Potential. 1. Adjoint
  (45) And Spinorial (16),}
  Phys.\ Rev.\  D {\bf 24}, 1005 (1981).
  %%CITATION = PHRVA,D24,1005;%%

%\bibitem{Yasue:1980qj}
  M.~Yasue,
  {\it How To Break SO(10) Via SO(4) X SO(6) Down To SU(2)(L) X SU(3)(C) X U(1),}
  Phys.\ Lett.\  B {\bf 103}, 33 (1981).
  %%CITATION = PHLTA,B103,33;%%

\bibitem{Anastaze:1983zk}
  G.~Anastaze, J.~P.~Derendinger and F.~Buccella,
  {\it Intermediate Symmetries In The SO(10) Model With (16+16) + 45 Higgses,}
  Z.\ Phys.\  C {\bf 20}, 269 (1983).
  %%CITATION = ZEPYA,C20,269;%%

\bibitem{Babu:1984mz}
  K.~S.~Babu and E.~Ma,
  {\it Symmetry Breaking In SO(10): Higgs Boson Structure,}
  Phys.\ Rev.\  D {\bf 31}, 2316 (1985).
  %%CITATION = PHRVA,D31,2316;%%

%bibitem{Harvey:1981hk}
%  J.~A.~Harvey, D.~B.~Reiss and P.~Ramond,
%  {\it Mass Relations And Neutrino Oscillations In An SO(10) Model,}
%  Nucl.\ Phys.\  B {\bf 199}, 223 (1982).
%  %%CITATION = NUPHA,B199,223;%%


%%%%%%%%-MOST GENERAL 45-16 POTENTIAL-%%%%%%%%%%%%%%%%%%%%%%%%%%%%%%%%%%%%%%%%%

\bibitem{Li:1973mq}
  L.~F.~Li,
  {\it Group Theory Of The Spontaneously Broken Gauge Symmetries,}
  Phys.\ Rev.\  D {\bf 9}, 1723 (1974).
  %%CITATION = PHRVA,D9,1723;%%

\bibitem{Buccella:1980qb}
  F.~Buccella, H.~Ruegg and C.~A.~Savoy,
  {\it Spontaneous Symmetry Breaking In O(10),}
  Phys.\ Lett.\  B {\bf 94}, 491 (1980).
  %%CITATION = PHLTA,B94,491;%%


%%%%%%%%-EFF TH MATCH-%%%%%%%%%%%%%%%%%%%%%%%%%%%%%%%%%%%%%%%%%%%%%%%%%%%%%%%%%%%

%bibitem{Weinberg:1980wa}
%  S.~Weinberg,
%  {\it Effective Gauge Theories,}
%  Phys.\ Lett.\  B {\bf 91}, 51 (1980).
%  %%CITATION = PHLTA,B91,51;%%

%bibitem{Hall:1980kf}
%  L.~J.~Hall,
%  {\it Grand Unification Of Effective Gauge Theories,}
%  Nucl.\ Phys.\  B {\bf 178}, 75 (1981).
%  %%CITATION = NUPHA,B178,75;%%


%%%%%%%%%%%%%%%% FLIPPED SU(5) %%%%%%%%%%%%%%%%%%%%%%%%%%%%%%%%%%%%

\bibitem{DeRujula:1980qc}
  A.~De Rujula, H.~Georgi and S.~L.~Glashow,
  {\it Flavor Goniometry By Proton Decay,}
  Phys.\ Rev.\ Lett.\  {\bf 45}, 413 (1980).
  %%CITATION = PRLTA,45,413;%%

\bibitem{Barr:1981qv}
  S.~M.~Barr,
  {\it A New Symmetry Breaking Pattern For SO(10) And Proton Decay,}
  Phys.\ Lett.\  B {\bf 112}, 219 (1982).
  %%CITATION = PHLTA,B112,219;%%


%%%%%%%% CW %%%%%%%%%%%%%%%%%%%%%%%%%%%%%%%%%%%%%%%%%%%%%%%%%%%%%%%%%

\bibitem{Coleman:1973jx}
  S.~R.~Coleman and E.~J.~Weinberg,
  {\it Radiative Corrections As The Origin Of Spontaneous Symmetry Breaking,}
  Phys.\ Rev.\  D {\bf 7}, 1888 (1973).
  %%CITATION = PHRVA,D7,1888;%%


%%%%%%%%-EXT SURVIVAL HYP-%%%%%%%%%%%%%%%%%%%%%%%%%%%%%%%%%%%%%%%%%

\bibitem{del Aguila:1980at}
  F.~del Aguila and L.~E.~Ibanez,
  {\it Higgs Bosons In SO(10) And Partial Unification,}
  Nucl.\ Phys.\  B {\bf 177}, 60 (1981).
  %%CITATION = NUPHA,B177,60;%%

%bibitem{Mohapatra:1982aq}
%  R.~N.~Mohapatra and G.~Senjanovic,
%  {\it Higgs Boson Effects In Grand Unified Theories,}
%  Phys.\ Rev.\  D {\bf 27}, 1601 (1983).
%  %%CITATION = PHRVA,D27,1601;%%


%%%%%%%%%%%%%%%% SO(10) GROUP THEORY %%%%%%%%%%%%%%%%%%%%%%%%%%%%%%%%%%%%

\bibitem{Slansky:1981yr}
  R.~Slansky,
  {\it Group Theory For Unified Model Building,}
  Phys.\ Rept.\  {\bf 79}, 1 (1981).
  %%CITATION = PRPLC,79,1;%%

\bibitem{Georgi:1982jb}
  H.~Georgi,
  {\it Lie Algebras In Particle Physics. From Isospin To Unified Theories,}
  Front.\ Phys.\  {\bf 54}, 1 (1982).
  %%CITATION = FRPHA,54,1;%%

%bibitem{Georgi:1979dq}
%  H.~Georgi and D.~V.~Nanopoulos,
%  {\it Ordinary Predictions From Grand Principles: T Quark Mass In O(10),}
%  Nucl.\ Phys.\  B {\bf 155}, 52 (1979).
%  %%CITATION = NUPHA,B155,52;%%

%bibitem{Fukuyama:2004ps}
%  T.~Fukuyama, A.~Ilakovac, T.~Kikuchi, S.~Meljanac and N.~Okada,
%  {\it SO(10) group theory for the unified model building,}
%  J.\ Math.\ Phys.\  {\bf 46}, 033505 (2005)
%  [arXiv:hep-ph/0405300].
%  %%CITATION = JMAPA,46,033505;%%



\end{thebibliography}
\end{document}